\documentclass[twocolumn]{aastex631}
\usepackage{amsmath,amsfonts,amssymb}
\usepackage{lipsum}

\newcommand{\UWM}{University of Wisconsin-Milwaukee, Milwaukee, WI 53201, USA}
\newcommand{\CMU}{McWilliams Center for Cosmology, Department of Physics, Carnegie Mellon University, Pittsburgh, PA 15213, USA}
\usepackage{appendix}

\begin{document}

\title{Searching for binary black hole sub-populations in gravitational wave data using binned Gaussian processes}

%\shorttitle{GPpop Subpopulations}

\author[0000-0002-7322-4748]{Anarya Ray}
\email{anarya@uwm.edu}
\affiliation{\UWM}
\author[0000-0003-2362-0459]{Ignacio Maga\~na~Hernandez}
\email{imhernan@andrew.cmu.edu}
\affiliation{\CMU}
\author[0000-0001-5228-6598]{Katelyn Breivik}
\affiliation{\CMU}
\author[0000-0003-3600-2406]{Jolien Creighton}
\affiliation{\UWM}

\shortauthors{Ray et al.}

\begin{abstract}

Astrophysically motivated population models for binary black hole observables are often insufficient to capture the imprints of multiple formation channels. This is mainly due to the strongly parametrized nature of such investigations. Using a non-parametric model for the joint population-level distributions of binary black hole component masses and effective inspiral spins, we find hints of multiple subpopulations in the third gravitational wave transient catalog. The higher (more positive) spin subpopulation is found to have a mass-spectrum without any feature at the $30-40M_{\odot}$, which is consistent with the predictions of isolated stellar binary evolution, simulations for which place the pile up due to pulsational pair-instability supernovae near $50M_{\odot}$ or higher. The other sub-population with effective spins closer to zero shows a feature at $30-40M_{\odot}$ and is consistent with binary black holes formed dynamically in globular clusters, which are expected to peak around $30M_{\odot}$. We also compute merger rates for these two subpopulations and find that they are consistent with the theoretical predictions of the corresponding formation channels. We validate our results by checking their robustness against variations of several model configurations and by analyzing large simulated catalogs with the same model.

\end{abstract}

\section{Introduction}

The formation of compact object binaries that merge within a Hubble time via gravitational wave~(GW) emission, particularly stellar-mass binary black holes~(BBHs), is poorly understood yet critically important to several fields of modern astrophysics~\citep{Mandel:2018hfr, Mapelli:2018uds, Mapelli:2020vfa}. The proposed formation scenarios for such BBH systems can be broadly classified into three categories: isolated evolution of massive stellar binaries that undergo orbital hardening, via either common-envelope, stable mass transfer, or chemical mixing~\citep[e.g., ][]{Postnov:2006hka,1976IAUS...73...75P,vandenHeuvel:2017pwp, Marchant:2016wow,Mandel:2015qlu}, dynamical assembly assisted by either a tertiary companion, multiple exchanges in dense clusters, or gas-assisted migration~\cite[e.g.,][]{Wen:2002km,Antonini:2012ad,Benacquista:2011kv,Bartos:2016dgn}, and hierarchical mergers in star clusters and AGN disks~\citep{Gerosa:2021mno}. These formation channels are usually characterized by several unknown parameters and initial conditions which can uniquely shape and correlate the observable features in an astrophysical BBH population, and hence be constrained from a catalog of GW observations.

At the end of its third observing run, the LIGO-Virgo-KAGRA detector network~\citep[LVK, ][]{LIGOScientific:2014pky,VIRGO:2014yos,KAGRA:2020agh} unveiled a catalog of GW observations~\citep[GWTC-3, ][]{KAGRA:2021vkt} around 70 of which were confidently\footnote{With a false alarm rate of less than 1 per year} identified to be BBH signals. Studying the ensemble properties of these BBHs has offered a plethora of new insights into the underlying astrophysics of their progenitor systems and environments. With the LVK's ongoing fourth observing run promising a large number of new detections, the search for the imprints of BBH formation mechanisms on the observable population properties of BBHs will continue to enable exciting new discoveries. Traditional approaches towards conducting such investigations have relied on the theoretical predictions of one or more formation models to construct a fiducial form of the distribution function of BBH parameters. The hyperparameters characterizing this distribution function encode information on the astrophysical processes underlying the corresponding formation scenarios and are inferred from a collection of GW observations~\citep[e.g.][]{Fishbach:2017zga,Edelman:2021fik,Kimball:2020qyd}. 

These \textit{parametric} population models, while useful in extracting information on a restrictive set of astrophysical assumptions regarding BBH formation, are not flexible enough to capture all of the physics that underlies BBH formation and are also susceptible to model-induced biases. For example, BBHs formed in isolation are expected to be distributed as a Power-law in component masses extending up to a maximum range followed by a steep fall-off known as the pair-instability gap and a small bump right before the truncation arising from the pile-up due to pulsational pair-instability supernovae~\citep[PPISN, ][]{Fishbach:2017zga, PPISN,plpp0}. However, when this \textsc{Powerlaw+Peak} model for the BBH mass distribution was constrained using GWTC-3 data~\citep{KAGRA:2021duu}, several concerns regarding our understanding of massive stellar interiors came to light. Firstly, the location of the PPISN feature was found to be within the $30-40M_{\odot}$ mass range, leading to tension with the predictions of stellar evolution simulations which place it above $50M_{\odot}$~\citep[see eg, ][ and references therein]{Hendriks:2023yrw}. Secondly, flexible alternatives to the \textsc{Powerlaw+Peak} model have found hints of additional features in the BBH mass spectrum~\citep{Edelman:2022ydv,Toubiana:2023egi, KAGRA:2021duu,Callister:2023tgi,Tiwari:2020vym}, which might be indicative of additional physics at play, beyond the restrictions built into the underlying assumptions of \textsc{Powerlaw+Peak}.

%A plausible explanation for both of these issues could be the existence of multiple sub-populations of BBHs in GWTC-3 data, each corresponding to different formation channels. The peak at the $30-40M_{\odot}$ range, for example, can also be explained by a sup-population of dynamically formed BBHs in globular clusters, and by isolated stellar binaries undergoing stable mass transfer

%which are expected to peak around the same mass range~\citep{Mapelli:2021gyv,Chattopadhyay:2023pil,Rodriguez:2018pss,Rodriguez:2021qhl, Wong:2020ise}. Given that the spin distribution of dynamically formed BBHs in globular clusters is expected to be more symmetric about zero than that of BBHs formed in isolation, observational evidence for these two sub-populations coexisting together can manifest in the form of population-level correlations between BBH masses and spins, and hence be searched for in current and future GW catalogs. However, several other astrophysical processes such as mass-ratio reversal in isolated binaries~\citep{Broekgaarden:2022nst} can introduce mass-spin correlations in the BBH population. Hence, the use of flexible population models that can account for a wide variety of known, as well as previously unmodelled astrophysical scenarios, is crucial to these kinds of investigations.  

A plausible explanation for both of these issues could be the existence of multiple sub-populations of BBHs in GWTC-3 data, each corresponding to different formation channels. The peak at the $30-40M_{\odot}$ range, for example, can be explained by a sub-population of dynamically formed BBHs in globular clusters, which are expected to peak around the same mass range~\citep{Antonini:2022vib, Wong:2020ise}. Given that the distribution of 
the effective inspiral spin parameters for dynamically formed BBHs in globular clusters is expected to be more symmetric about zero than that of BBHs formed in isolation~\citep{Mapelli:2021gyv,Chattopadhyay:2023pil,Rodriguez:2018pss,Rodriguez:2021qhl}, observational evidence for these two sub-populations coexisting together can manifest in the form of population-level correlations between BBH masses and spins, and hence be searched for in current and future GW catalogs~\citep{Baibhav:2022qxm}. However, several other astrophysical processes such as mass-ratio reversal in isolated binaries~\citep{Broekgaarden:2022nst} can introduce mass-spin correlations in the BBH population. Hence, the use of flexible population models that can account for a wide variety of known, as well as previously unmodelled astrophysical scenarios, is crucial to these kinds of investigations.  

In this letter, using a non-parametric model for the joint population-level distribution of BBH component masses and effective inspiral spins based on binned Gaussian processes~\citep[BGP, ][]{Ray:2023upk,Mohite:2022pui,Mandel:2016prl}, we find hints of multiple subpopulations of BBHs in GWTC-3 data, one consistent with the predictions of isolated BBH formation, and the other dynamical. % I don't know if we should say consistent with isolated BBH formation here. 
In particular, we find two distinct shapes in the BBH mass spectrum, one of which has a powerlaw-like fall-off with no feature in the $30-40M_{\odot}$ region and is associated with higher (more positive) values of effective inspiral spins. The other one is found to have a bump at the $30-40M_{\odot}$ and is associated with smaller values of effective inspiral spins. Furthermore, BBHs with at least one component in the $30-40M_{\odot}$ range are found to have a more symmetric distribution of effective inspiral spins about zero, as compared to that of the complementary case which is characterized by a spin distribution skewed towards more positive values. We also constrain the total merger rate of BBHs within different mass and effective spin ranges and find that they are consistent with the theoretical predictions of the corresponding formation models.
%need to check this too.

Given the size of contemporary datasets, we restrict the dimensionality of our population model to account for the distributions of only the best-measured among BBH observables that are relevant to the astrophysics we aim to extract. In particular, we only model the joint distribution of source frame component masses and effective inspiral spins using BGPs, and fix that of other BBH observables to be of much simpler and less flexible functional forms. While this restriction amounts to ignoring higher dimensional correlations in the BBH population that can in principle bias our conclusions, we validate our analysis using large simulated catalogs and demonstrate that the ignored correlations cannot manifest into the trends we find in the joint mass and effective spin distributions. Even though a higher dimensional BGP on masses, redshifts, and all of the spin parameters (component spin magnitudes and tilts) would be the most generalized and robust approach to this investigation, we expect such a model to yield uninformative constraints given GWTC-3-sized datasets. However, as catalogs continue to grow, we aim to implement higher dimensional BGP models for obtaining a more generic and unbiased characterization of the various sub-populations of BBHs.

In a previous study, \cite{Godfrey:2023oxb} had also looked at mass-spin correlations in the BBH population using a mixture of semi-parametric models and had come to conclusions that are broadly consistent with ours except for a few subtle distinctions. They identify a sub-population of binaries with a mass distribution that demonstrates a strong peak near $10M_{\odot}$ followed by a sharp fall-off, and a spin distribution that slightly prefers aligned components over isotropic ones. They identify a second sub-population that demonstrates a peak in the 30-40$M_{\odot}$ range and has a spin distribution consistent with that of the other subpopulation. In addition to component spins, they infer an effective spin distribution that is completely consistent between the two subpopulations they identify. On the other hand, our study identifies two subpopulations that correspond to significantly different effective spin distributions. The astrophysical implications of \cite{Godfrey:2023oxb} are therefore slightly different from ours even though both are broadly in agreement regarding the existence of a subpopulation in the mass-spectrum that falls off steadily beyond 15$M_{\odot}$. These differences might be attributed to the differences in our modeling choices which are discussed further in Sec.~\ref{sec:methods}.

This letter is organized as follows. In Sec.~\ref{sec:methods} we describe in detail our flexible population model and the associated inference framework. In Sec.~\ref{sec:results}, we summarize our results for GWTC-3 and describe the data used to obtain it. In Sec.~\ref{sec:validation}, we validate our model by analyzing large simulated catalogs of five different BBH populations and demonstrate that our results for real data are robust against known sources of systematic biases. In Sec.~\ref{sec:astro}, we discuss the astrophysical implications of our findings. Lastly, In Sec.~\ref{sec:conclusion}, we conclude with a summary of our model, findings, and future investigations.

\section{Population model and hierarchical inference}
\label{sec:methods}
To search for multiple sub-populations in the BBH mass spectrum that potentially correspond to different distributions of effective inspiral spin, we construct a flexible model for the population properties of BBH observables, agnostic of astrophysical predictions. Specifically, we model the joint distribution of BBH component masses~$(m_1,m_2)$, and effective inspiral spins~$(\chi_{\text{eff}})$ as a piece-wise binned function, such as:

\begin{widetext}
\begin{equation}
    \frac{dN}{dm_1dm_2dzd\chi_{\rm{eff}}}(m_1,m_2,z,\chi_{\rm{eff}}|\vec{n},\kappa)=\sum_{\gamma}\frac{n^{\gamma}}{m_1m_2}\frac{dV}{dz}T_{\rm{obs}}(1+z)^{\kappa-1}\times\begin{cases} 1 & \textrm{if }(m_1,m_2,\chi_{\rm{eff}})\in\gamma^{th}\textrm{ bin}\\0 & \textrm{otherwise}\end{cases}\label{pop-model}
\end{equation}
\end{widetext}
where, the LHS is the number of CBCs per component masses, effective spin and redshift, $n^{\gamma}$ is the merger rate density per comoving volume, source-frame time, log component mass, and effective spin in the $\gamma^{th}$ bin, and $\kappa$ is the parameter that controls the redshift evolution of the overall merger rate. Simultaneously inferring the merger rate density in each bin from multiple uncertain measurements amounts to learning the shape of the population distribution from GW data up to the resolution limit imposed by our choice of binning.

While Eq.~\eqref{pop-model} is flexible enough to capture any correlation between the mass and effective spin populations, it is unable to inform on the existence of mass-redshift and spin-redshift correlations. Even though contemporary datasets are mostly uninformative on the existence of mass-redshift correlations~\citep{KAGRA:2021duu}, particularly in the context of binned models~\citep{Ray:2023upk}, the studies of \cite{Heinzel:2023hlb} and \cite{Biscoveanu:2022qac}  have revealed strong evidence for spin-redshift correlations in the same. \cite{Biscoveanu:2022qac} has further shown that certain population models designed to search only for a mass-spin correlation can falsely infer their existence from a simulated catalog with spin-redshift correlations. We show is Sec.~\ref{sec:validation} that our results are robust against such model-induced systematics by testing our model on large simulated catalogs characterized by various correlated populations. See \cite{Rinaldi:2023bbd,Karathanasis:2022rtr} for more investigations on mass-redshift correlations using alternative models.% Point to section 4 here?

To infer the merger rate density in each bin from multiple uncertain measurements of BBH masses and effective spins we implement the framework of Bayesian hierarchical inference~\citep{Thrane:2018qnx,Mandel:2018mve,popgw2,popgw3}. By modeling the occurrence of BBHs as an inhomogeneous Poisson process, we construct the likelihood function of the rate densities from a collection of single event measurements and use a simulated population of detectable BBHs to account for Malmquist biases in the inferred distributions. In the context of our population model, the likelihood function takes the following form:

\begin{equation}
    \log p(\vec{d}|\vec{n})=-\sum_{\gamma}n^{\gamma}\left<VT\right>^{\gamma}+\sum_i\log\left(\sum_{\gamma}n^{\gamma}w^{\gamma}(d_i)\right)\label{likelihood}
\end{equation}
where $w^{\gamma}(d_i)$ is the posterior support of the $\gamma^{th}$ bin from the $i^{th}$ event and $\left<VT\right>^{\gamma}$ is the sensitive time-volume within which BBHs in the $\gamma^{th}$ bin are expected to be detectable~\citep{Ray:2023upk}. %For a discussion regarding the convergence of the Monte Carlo sums used to compute these quantities~\citep{Pdet1-Farr,Pdet2-essick,Ray:2023upk}, see appendix .%The posetrior weights and sensitive time-volumes are obtained by re-weighting the posterior distribution of BBH observables obtained form Bayesian parameter estimation of each event

To regularize the inferred shape of the population distribution over sparse regions of parameter space we choose the prior on logarithmic rate densities to be a Gaussian process~(GP) with an exponential quadratic kernel. The means, correlation lengths, and covariance amplitudes characterizing the GP are themselves modeled using normal, log-normal, and half-normal priors respectively~\citep{Ray:2023upk}. Using Hamiltonian Monte Carlo~(HMC) sampling implemented through the No U-Turn Sampler~\citep{HMC,HMC-NUTS}, we infer the GP hyper-parameters along with the rate densities characterizing our population model from their joint posterior distribution. The code developed to implement this analysis is publicly available as the python package \texttt{gppop}\footnote{https://github.com/AnaryaRay1/gppop/tree/spin-dev}, which in turn relies on the \texttt{Pymc} package~\citep{pymc2023} for conducting the HMC sampling.% given GW data from a catalog of BBH observations.%, which takes the following form:

Previous studies by \cite{Pdet1-Farr,Pdet2-essick} have demonstrated that the convergence of the Monte Carlo sums used to compute the posterior weights and detectable time-volumes of Eq.~\eqref{likelihood} needs to be ensured in order to avoid biases in the corresponding hierarchical inference. Following the implementation of their convergence criteria in the context of the binned model, as derived by~\cite{Ray:2023upk}, we verify that all of our inferred hyper-parameter samples support regions of parameter space that have enough effectively independent Monte Carlo samples, both for event-specific posteriors and detectable injections to avoid biases resulting from improperly converged estimates of posterior weights and detectable time-volumes.

Even though we do not fit for the redshift evolution parameter $\kappa$ and instead fix it to be $\kappa=2.9$, we show in appendix~\ref{sec:kappa} that varying $\kappa$ across its measured confidence interval leaves our results unchanged. The reason for this choice is as follows. For fixed $\kappa$, the posterior weights and sensitive time-volumes of Eq.~\eqref{likelihood} are pre-computable leading to computationally cheap inference of the joint population posterior distribution. Inferring $\kappa$ amounts to re-computing the weights and time-volumes for every step of the sampling which is intractable for our current Central Processing Unit~(CPU) based implementation. While parallelizing these computations on Graphics Processing Units~(GPUs) offers a potential solution to this problem, the associated code developments are part of an ongoing study and beyond the scope of this work given the insensitivity of our conclusions to the value of $\kappa$. We however note that our current implementation is much more scalable on CPUs than the earlier ones of \cite{Ray:2023upk}. See appendix~\ref{sec:scalability} for more details.

We further note that while previous attempts~\citep{Godfrey:2023oxb,Li:2023yyt} at obtaining data-driven constraints on mass-spin correlations in the BBH population have accounted for more spin parameters such as component spin magnitudes and orientations, most of these studies have imposed several restrictions on the nature of mass-spin correlations, in contrast to our flexible model. In particular, they have used mixture models wherein, each mixture component is assumed to have an uncorrelated mass-spin distribution. Therefore, within their framework, mass-spin correlations can only manifest in the form of multiple sub-populations coexisting together. Hence, mass-spin correlations arising from the astrophysical processes within a single formation channel \citep[such as mass-ratio reversal during stable mass transfer introducing a spin-mass-ratio anti-correlation during field formation][]{Broekgaarden:2022nst}, can potentially bias the astrophysical conclusions of such models. On the other hand, we impose no similar restrictions on our joint mass-effective-spin distribution and reconstruct its functional form directly from the data, up to the resolution limit imposed by our choice of binning.

An additional source of systematics in our population model involves the potential sensitivity of the inferred shapes of the population distribution to binning choices. Similar to the example presented in \cite{Ray:2023upk} we show in appendix~\ref{sec:binning} that our results are stable against multiple binning choices by varying the mass and $\chi_{\rm{eff}}$ bin locations and demonstrating that the trends we infer are robust against this variation.%\textcolor{orange}{AR to add text, either here or in appendix, about how increasing the number of bins is expected to broaden the posteriors given the size of contemporary datasets}%doubling the number of bins in the analysis of simulated data and demonstrating how the inferred correlation lengths of the GP remain unchanged leading to accurate recovery of the injected population distribution irrespective of binning choices. 
\section{Results}
We re-analyze public LVK data~\citep{gwosco3} comprising all the BBH events that were observed through GWTC-3 with a FAR of less than 1 per year. Following previous works, we exclude the \textit{outlier} event GW190814 given the uncertainty regarding its system of origin, which leaves us with a set of 69 high confidence BBH observations~\citep{LIGOScientific:2020kqk,LIGOScientific:2020zkf,KAGRA:2021duu,no1908141}. For each of these events, we use $(m_1,m_2,\chi_{\rm{eff}})$ samples publicly released by the LVK~\citep{gwosco3} to compute the posterior weights in Eq.~\eqref{likelihood}. Specifically, following~\citep{KAGRA:2021duu}, we convert detector frame mass and luminosity distance samples to source frame by assuming a particular cosmological model, which we choose to be Planck 2015~\citep{Planck:2015fie}. For further details of the single-event PE analyses see~\citep{KAGRA:2021duu}. We also use LVK's publicly released set of detectable injections~\citep{gwosco3} to compute the detectable time-volumes of Eq.~\eqref{likelihood}.
\label{sec:results}
\begin{figure*}[htt]
\begin{center}
\includegraphics[width=0.46\textwidth]{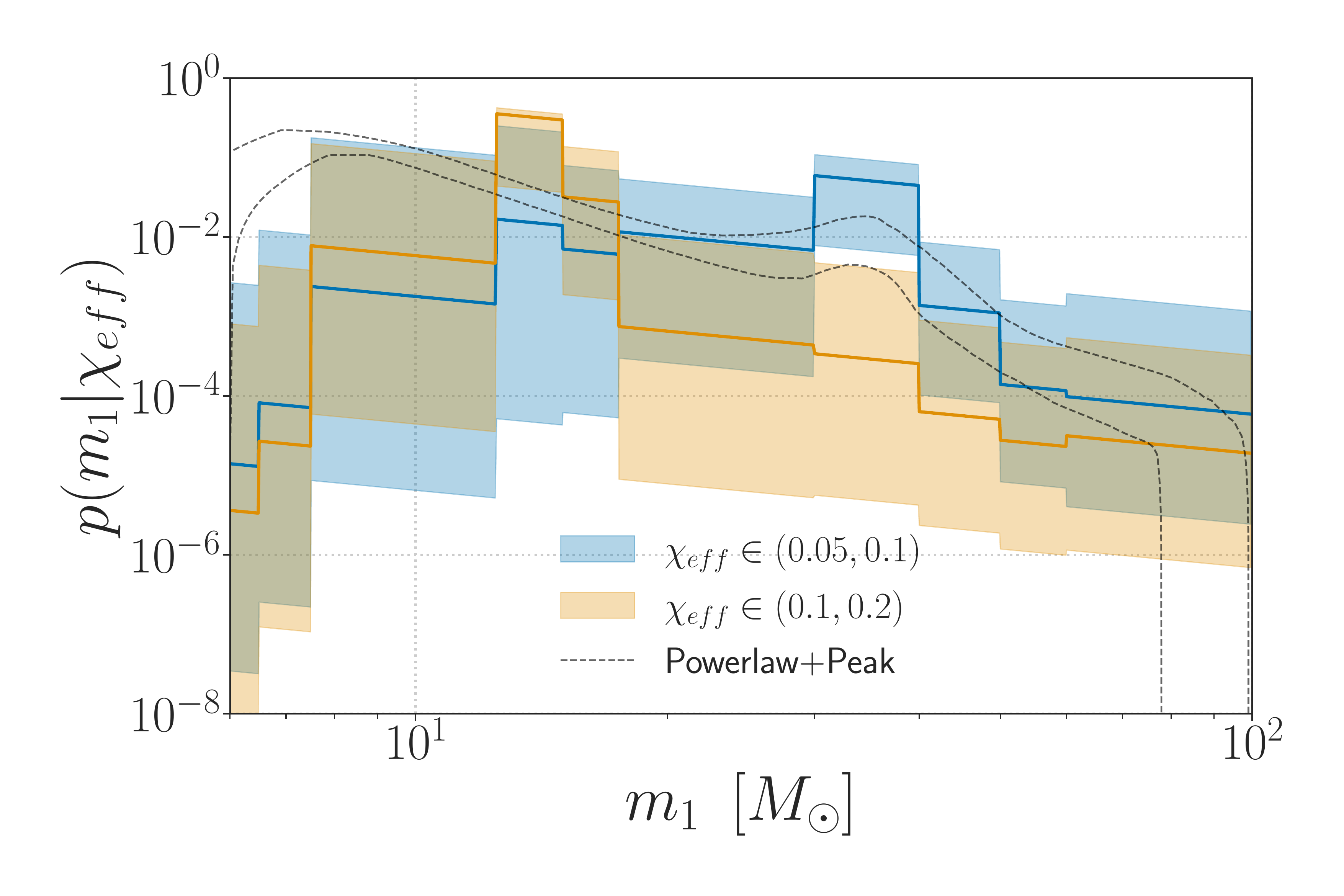}
\includegraphics[width=0.46\textwidth]{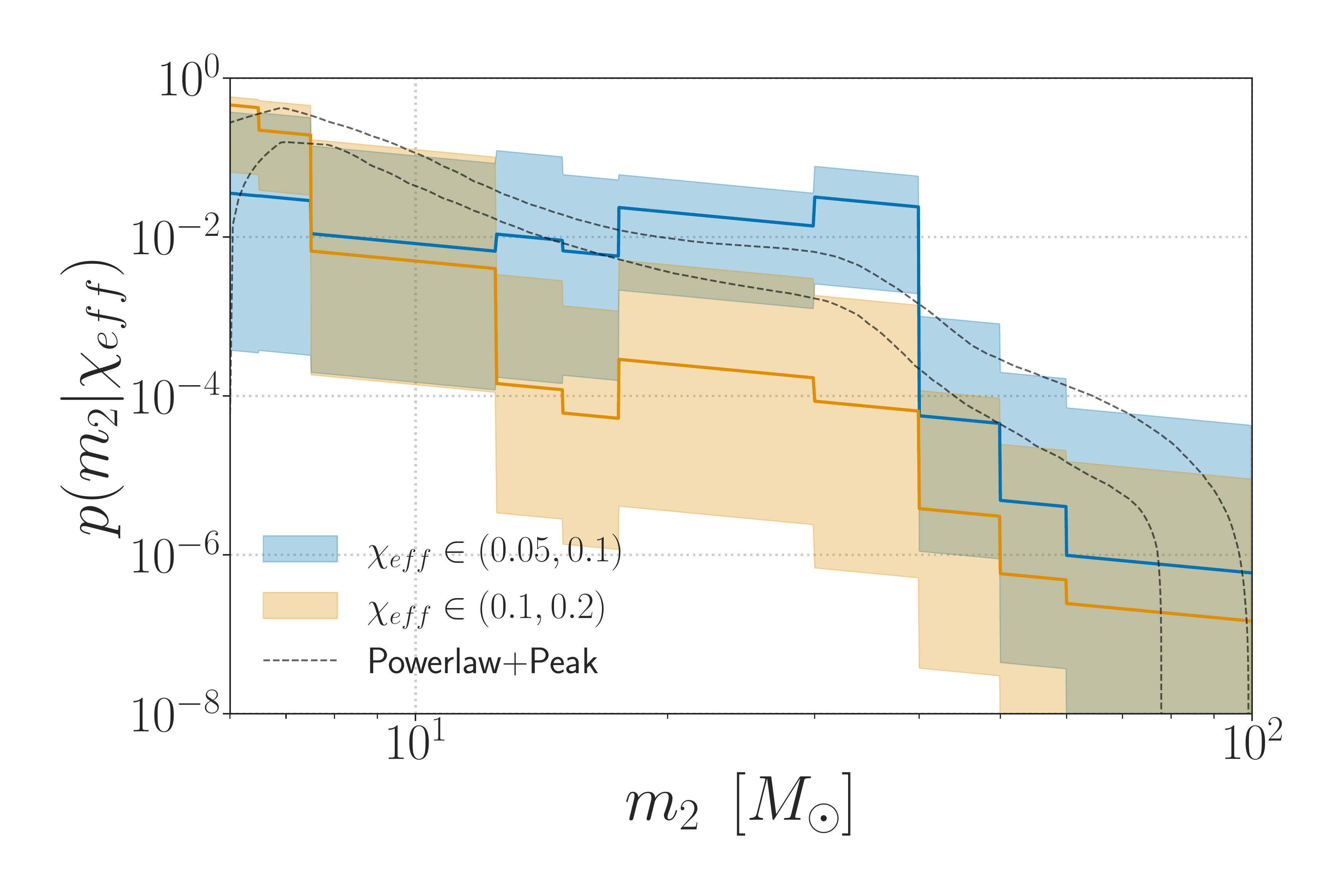}
\includegraphics[width=0.46\textwidth]{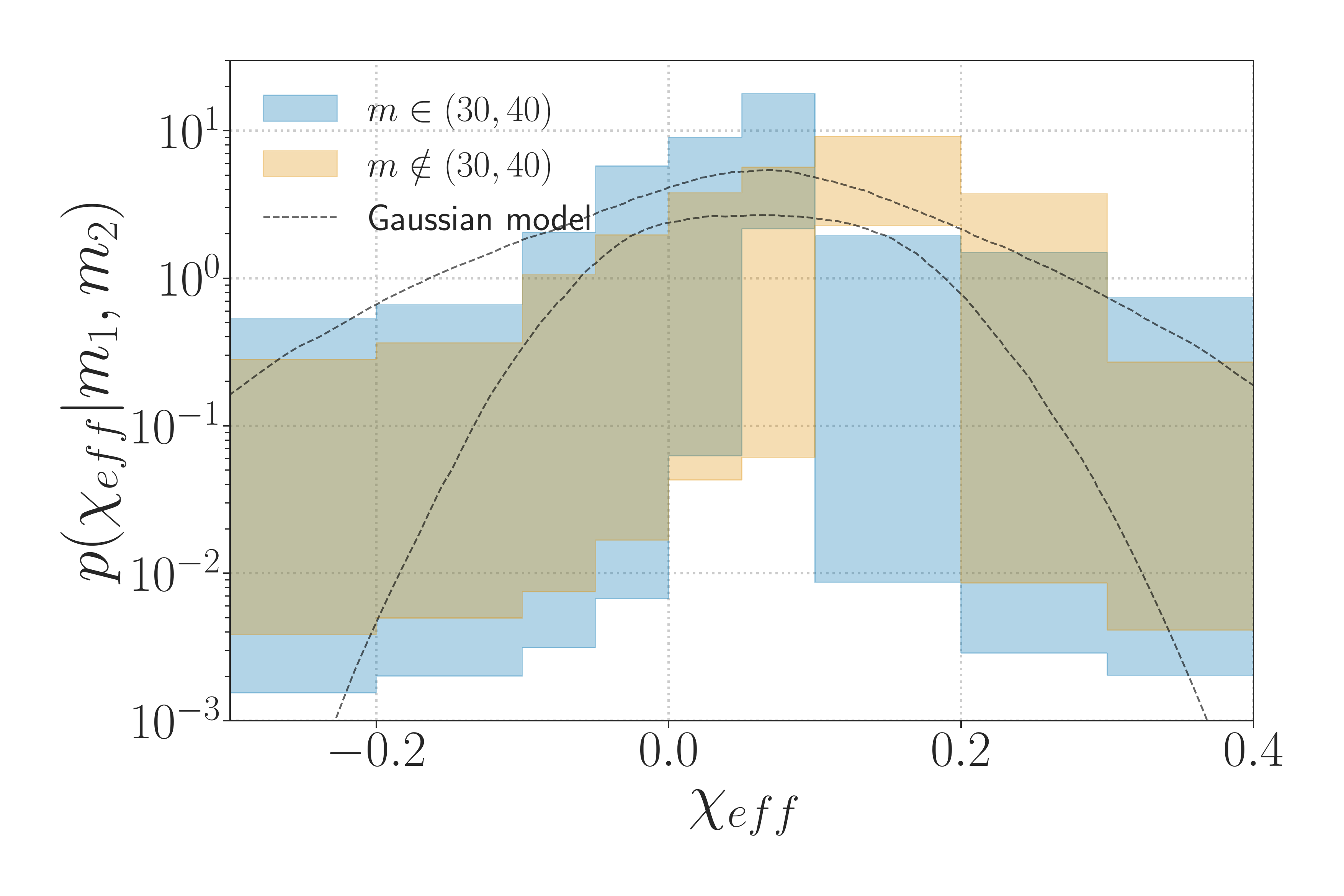}
\includegraphics[width=0.46\textwidth]{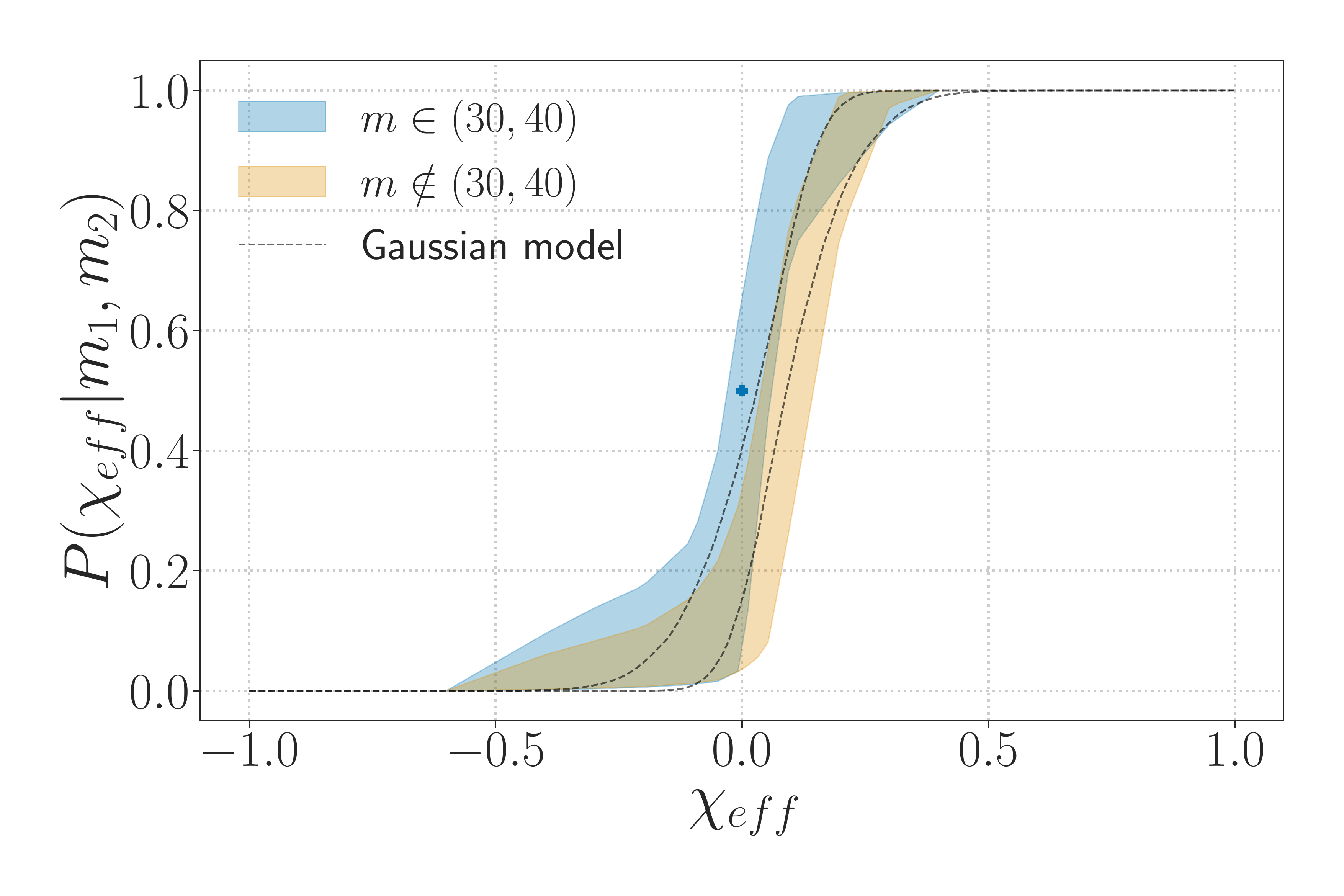}
\caption{\label{fig:GWTC-3-chi-q} Conditional distributions of BBH component masses and spins obtained from GWTC-3. The parametric inferences of \cite{KAGRA:2021duu} that used the Powerlaw+Peak and Gaussian effective spin models are overplotted for comparison. The top panels show the primary (\textit{left}) and secondary~(\textit{right}) mass distributions conditioned on two different ranges of effective inspiral spins, namely  $0.05<\chi_{\rm{eff}}<0.1$, and $0.1<\chi_{\rm{eff}}<0.2$. The bottom panels show the density~(\textit{left}) and cumulative densitiy~(\textit{right}) functions corresponding to the distribution of effective inspiral spins, conditioned on two different mass ranges, namely $\{m_1\in(30M_{\odot},40M_{\odot})\text{ or }m_2\in(30M_{\odot},40M_{\odot})\}$, and its complement. The blue dot on the lower right figure is the point (0,0.5). %KB: every panel should have it's own decription and potentially punchline depending on how long the caption gets. 
}
\end{center}
\end{figure*}

% KB: I think you should state clearly what the trends are just to be extra clear here.
As seen in Figure~\ref{fig:GWTC-3-chi-q}, we find trends indicative of at least two subpopulations in the BBH mass-spectrum that differ by 90\% in shape for the $30-40M_{\odot}$ bin and are each conditional on different ranges of effective inspiral spin. In particular, we find that for $0.1<\chi_{\rm{eff}}<0.2$, the primary and secondary mass distributions resemble a power-law like fall off with no feature in the $30-40M_{\odot}$ range. On the other hand, for $0.05<\chi_{\rm{eff}}<0.1$, we find that the primary and secondary mass-distributions both show an excess of BBHs in the $30-40M_{\odot}$ range. Similarly, constraints on the conditional spin distribution depict that binaries with one or both components in the $30-40M_{\odot}$ mass range have a more symmetric effective spin distribution about zero as compared to other binaries which are found to have a skewed distribution of $\chi_{\rm{eff}}$ that prefers more positive values. We also compute the fraction of events with $\chi_{\text{eff}}<0$ to be $21^{+43}_{-18}\%$ for binaries with at least one component in the the $30-40M_{\odot}$ mass range and $10^{+22}_{-7}\%$ for all other binaries. %We also find the combined merger rate of binaries in the  $30-40M_{\odot}$ mass range to be $2.04_{-0.73}^{+0.79} \ \rm{Gpc^{-3}yr^{-1}}$, and that of other binaries to be  $22.4_{-9.4}^{+10.2} \ \rm{Gpc^{-3}yr^{-1}}$. Similarly,  we find the combined merger rate of binaries in the  $0.05-0.1$ and $0.1-0.2$ spin ranges\footnote{As in ranges of effective inspiral spin} to be $2.1_{-1.8}^{+3.1} \ \rm{Gpc^{-3}yr^{-1}}$, and $16.1_{-11.5}^{+11.9} \ \rm{Gpc^{-3}yr^{-1}}$ respectively. We find the combined merger rate of BBHs to be $24.4_{-9.2}^{+10.7} \ \rm{Gpc^{-3}yr^{-1}}$.

\section{Validation Study}
\label{sec:validation}
\begin{figure*}[htt]
\begin{center}
\includegraphics[width=0.32\textwidth]{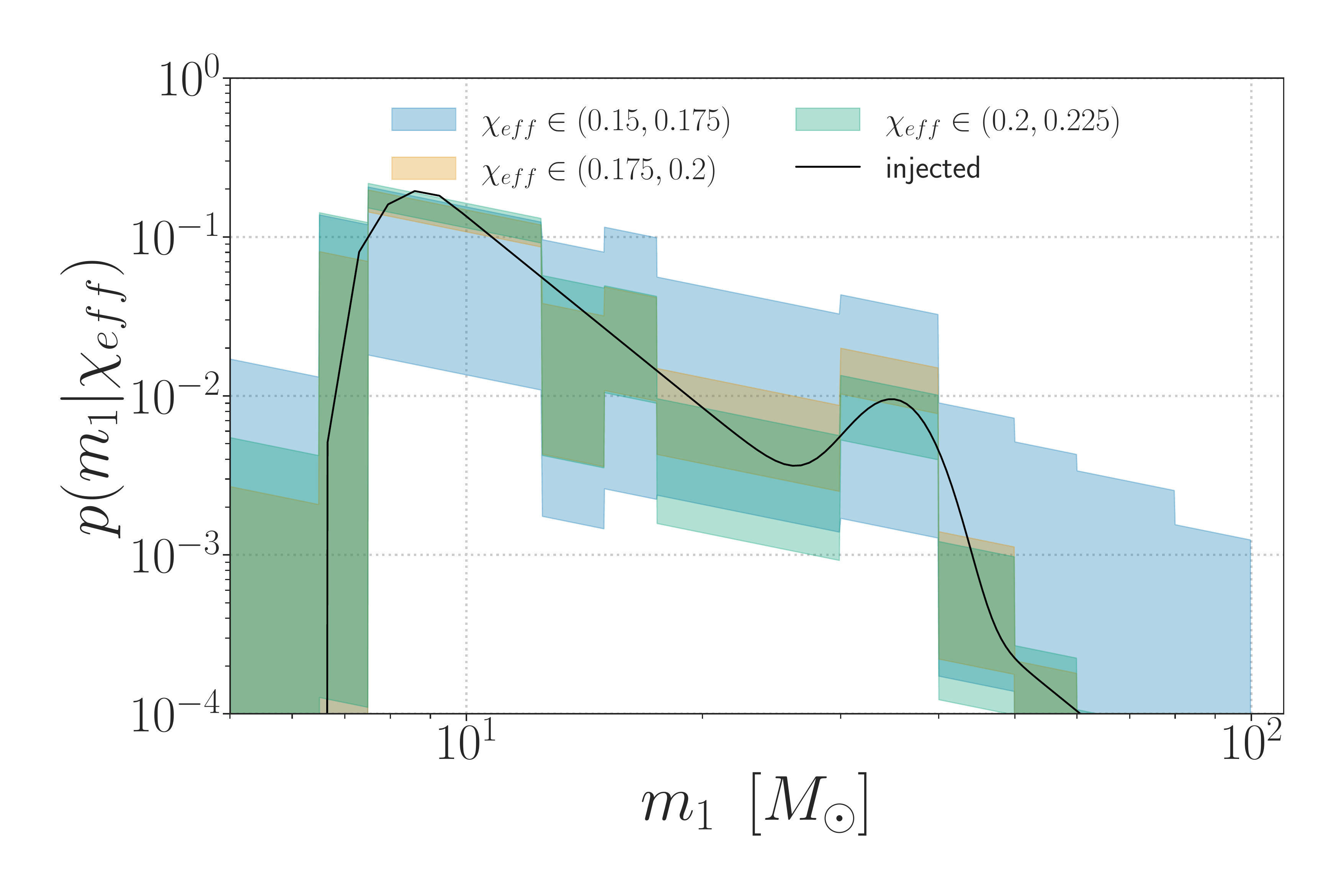}
\includegraphics[width=0.32\textwidth]{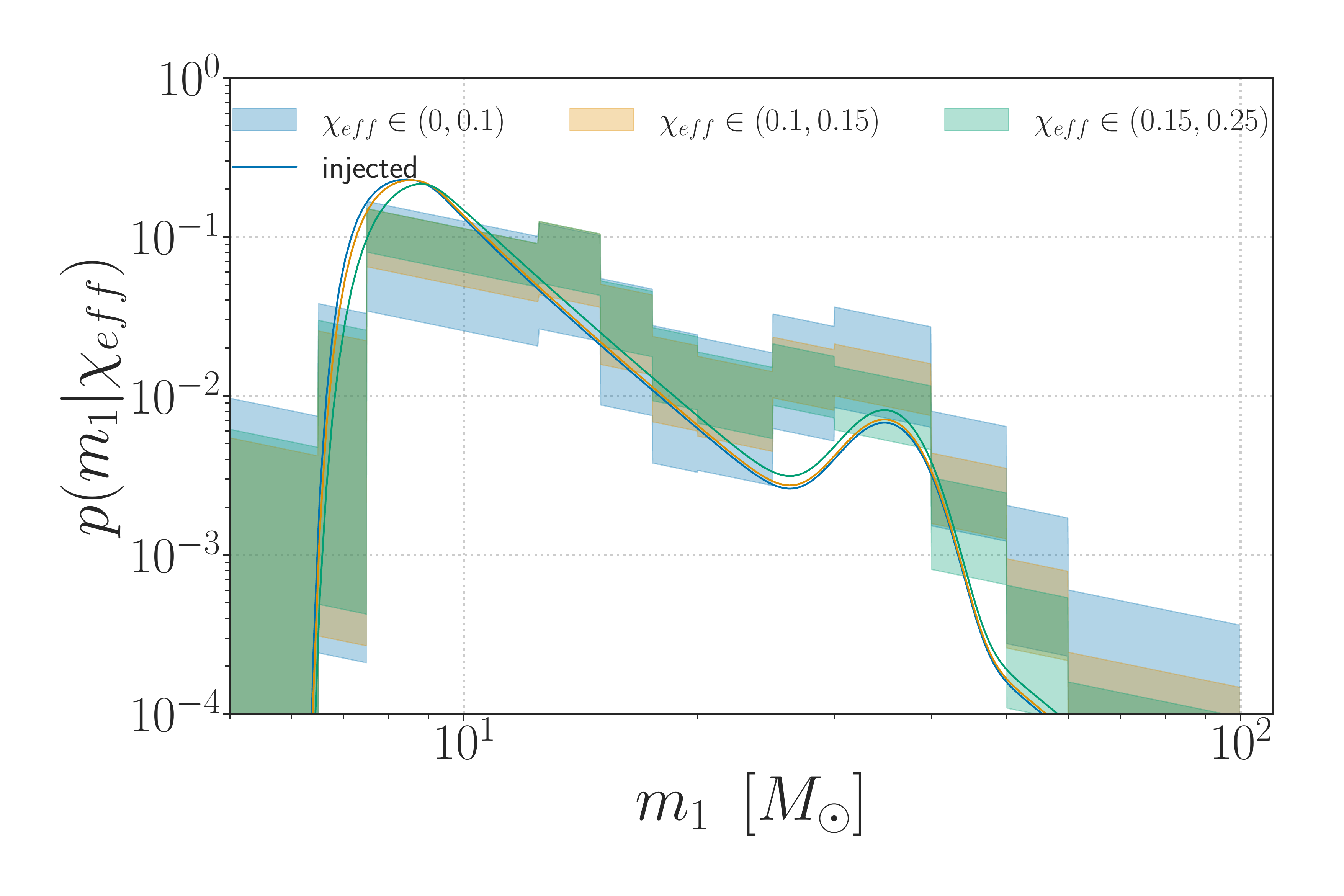}
\includegraphics[width=0.32\textwidth]{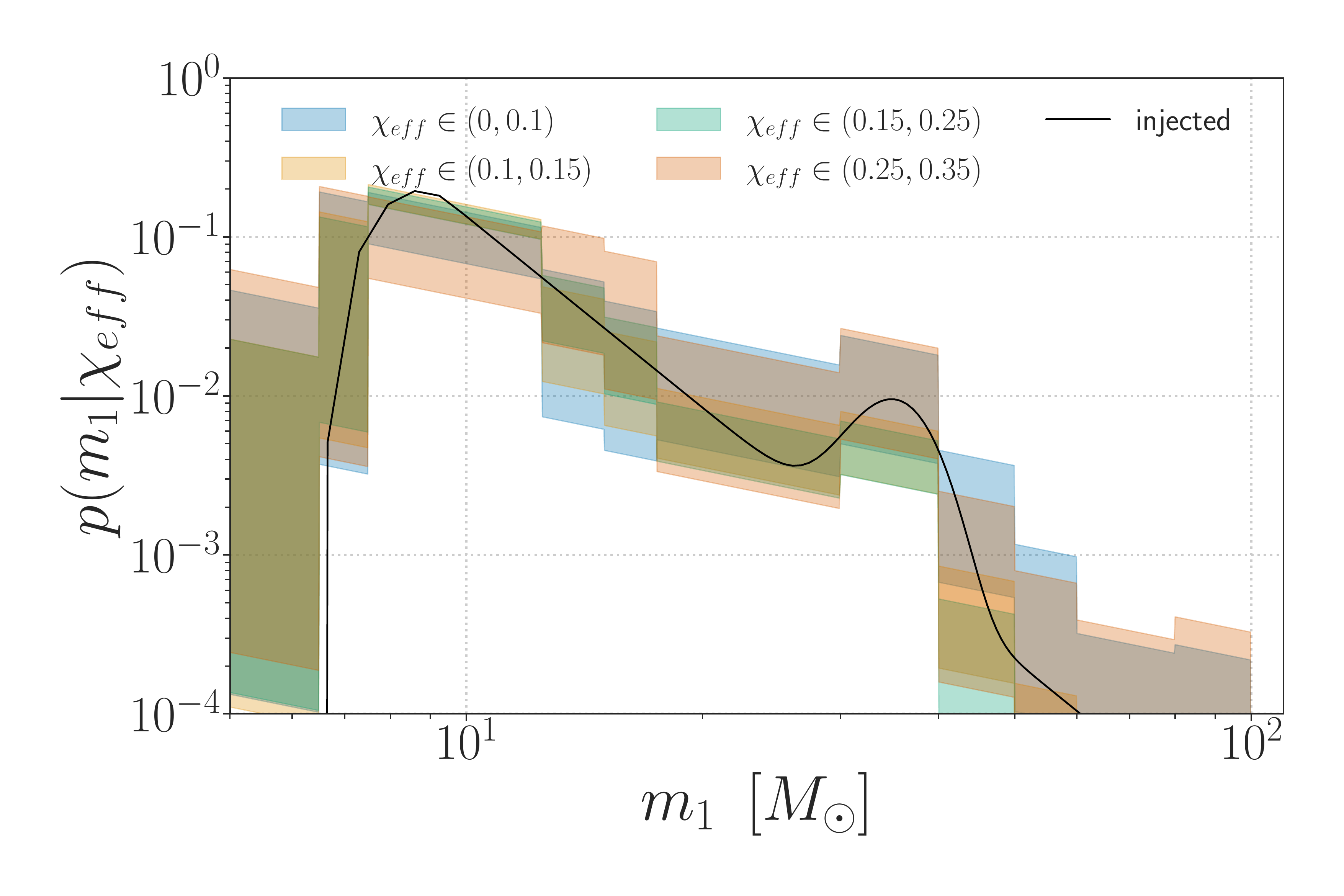}
\includegraphics[width=0.32\textwidth]{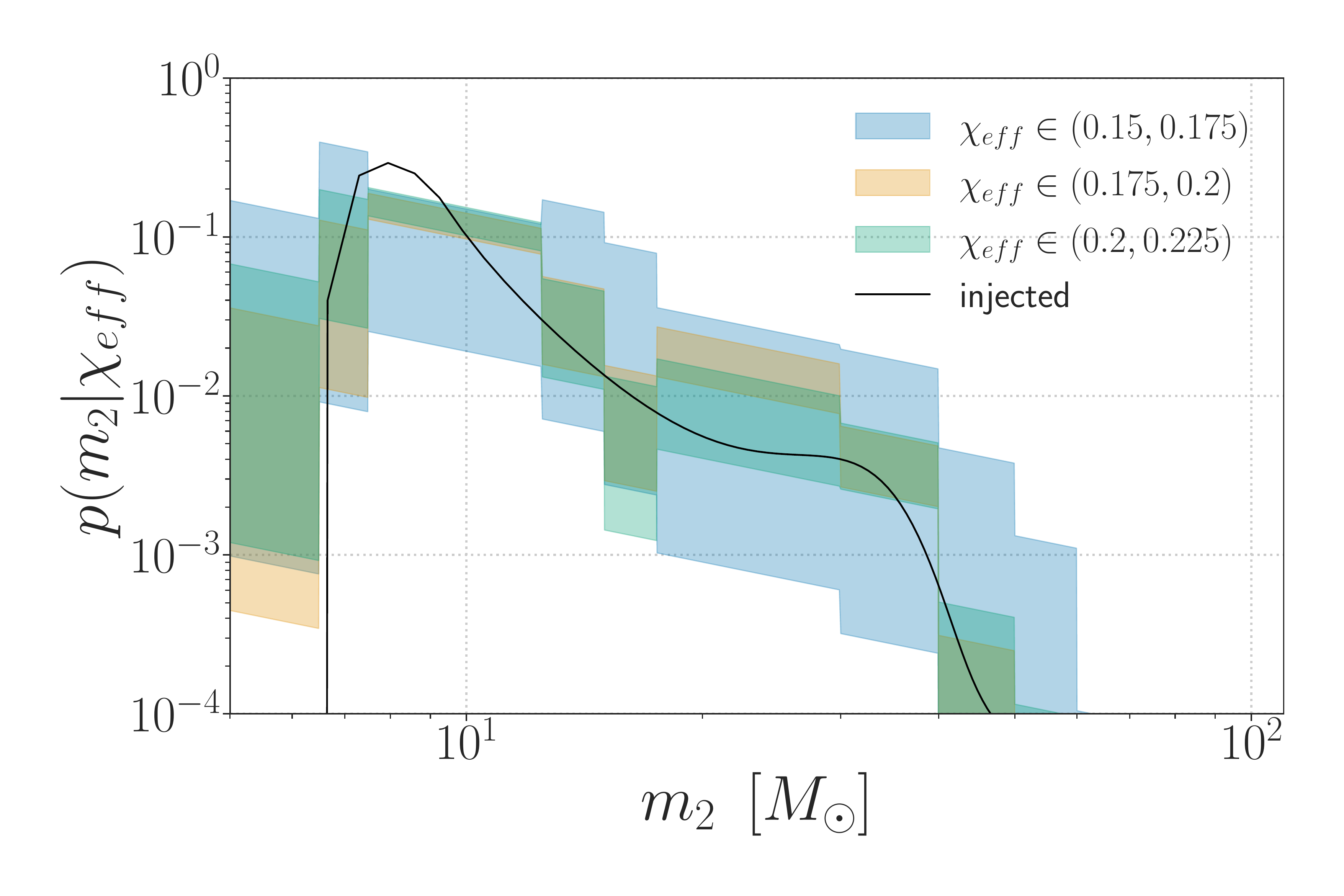}
\includegraphics[width=0.32\textwidth]{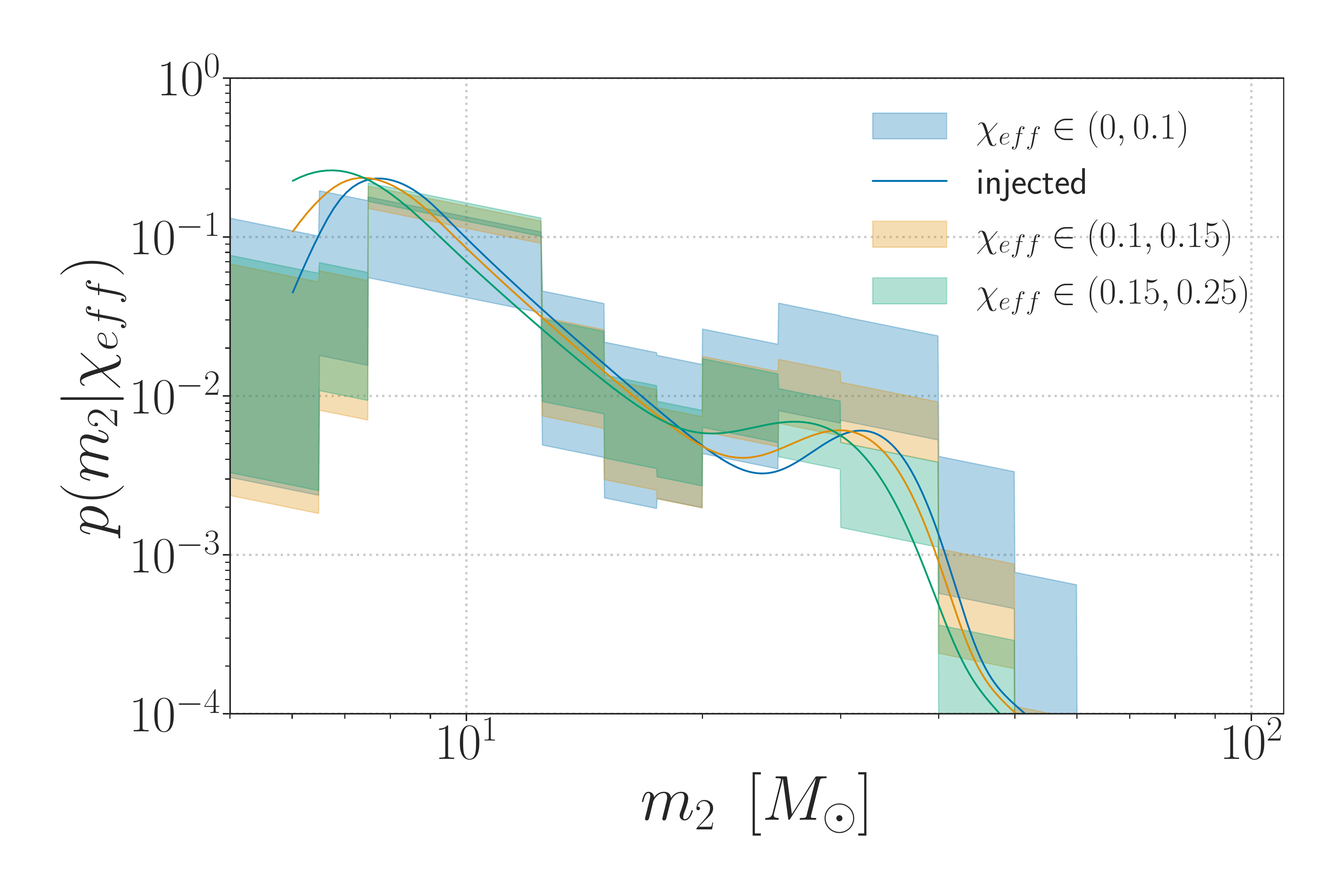}
\includegraphics[width=0.32\textwidth]{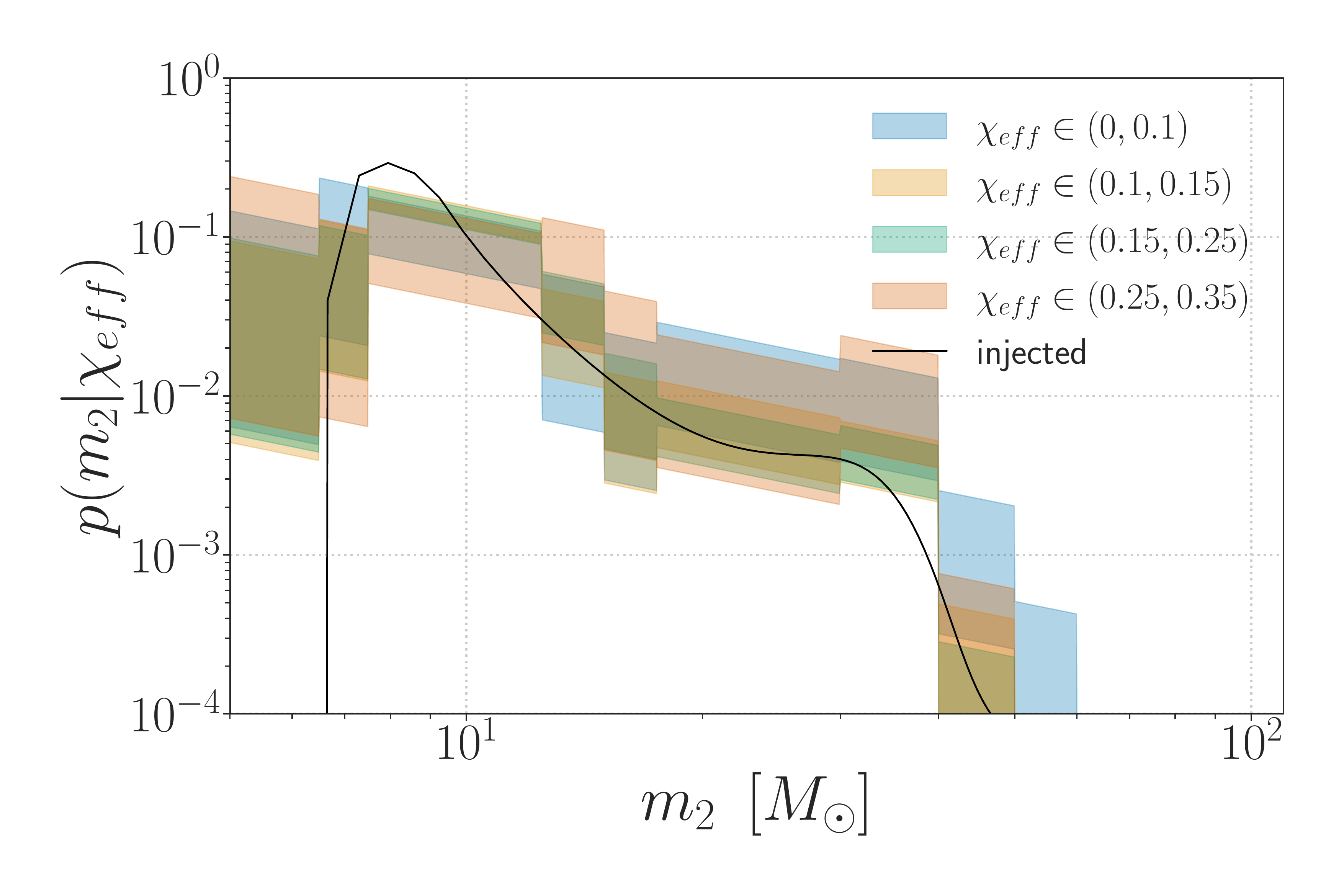}
\includegraphics[width=0.32\textwidth]{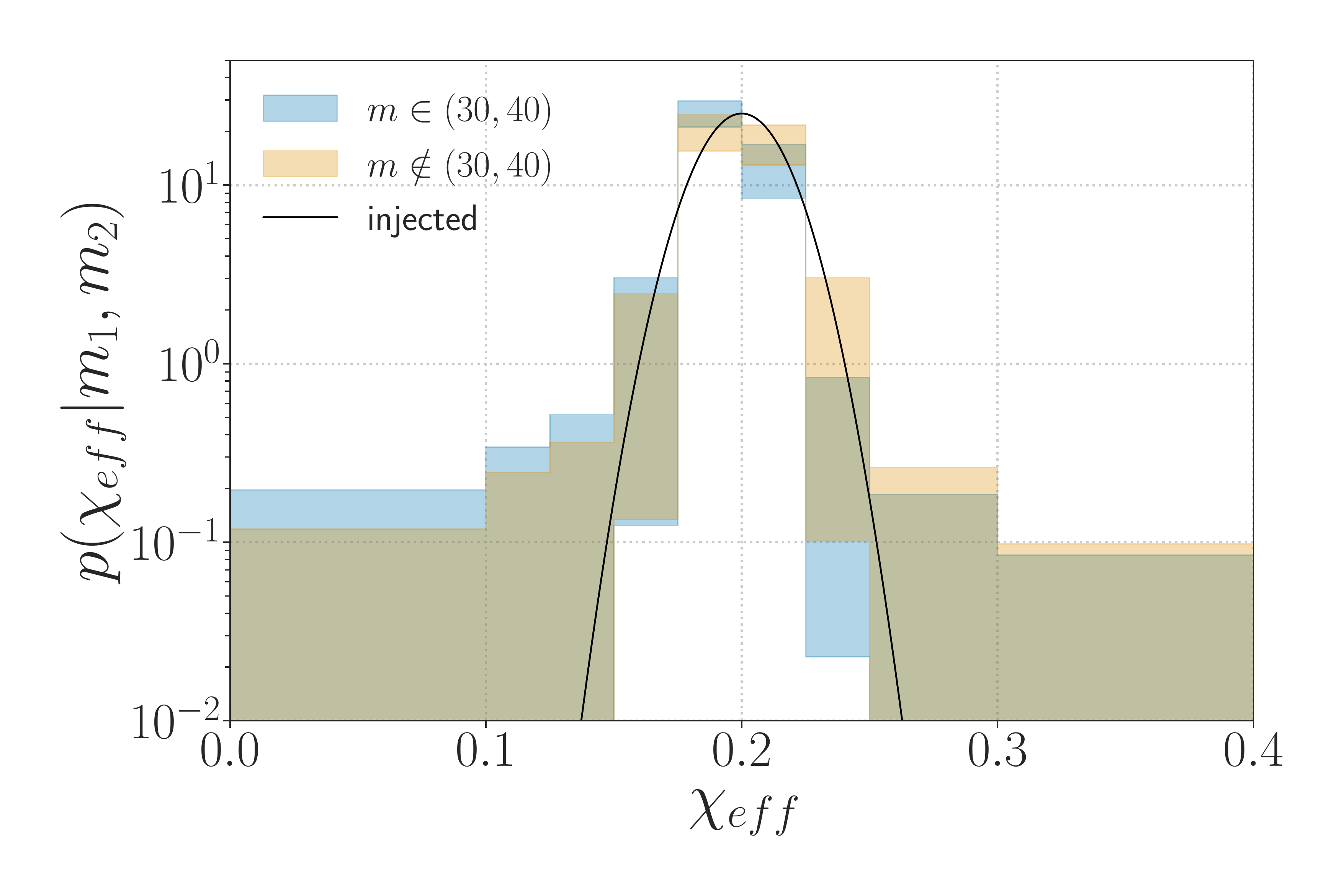}
\includegraphics[width=0.32\textwidth]{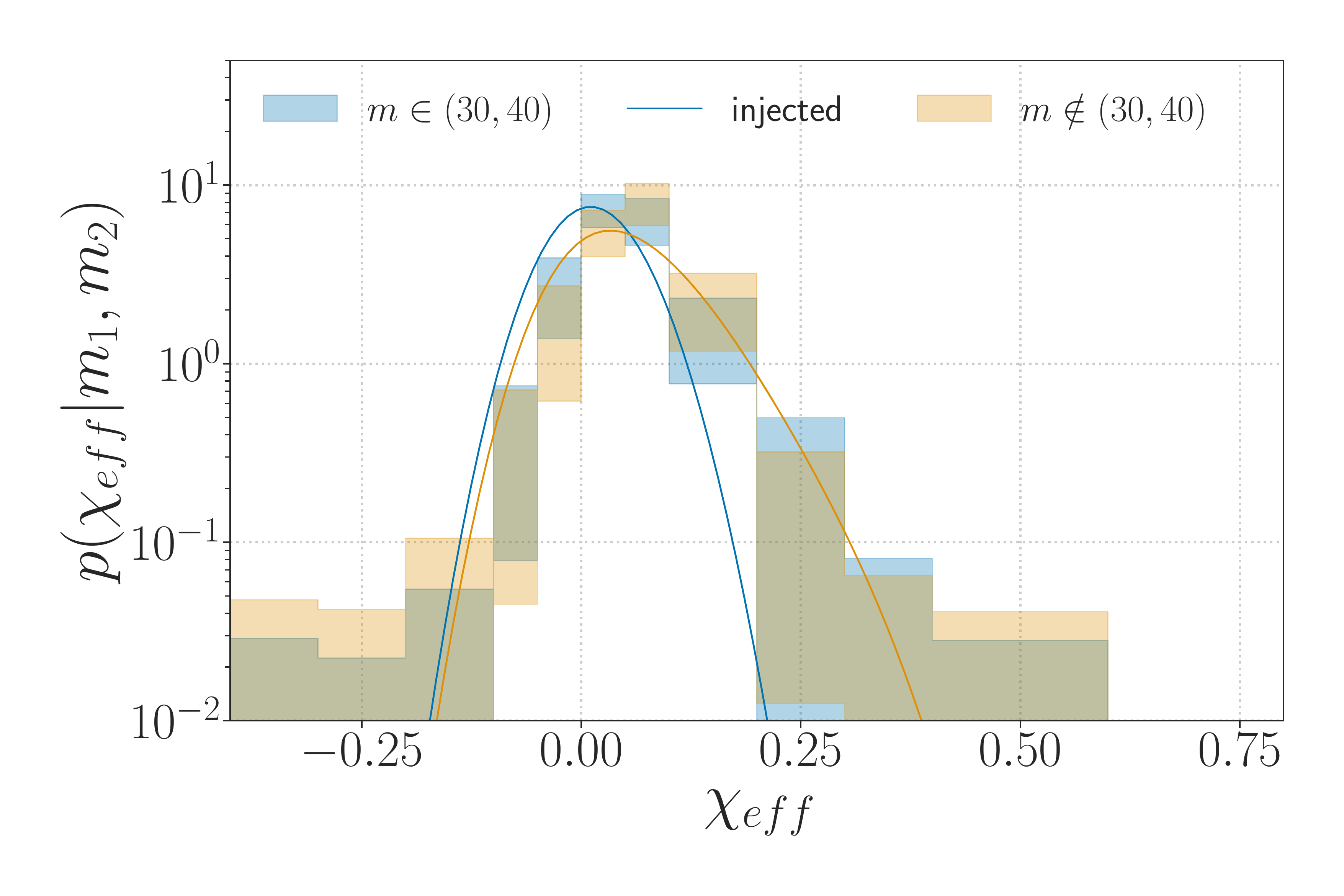}
\includegraphics[width=0.32\textwidth]{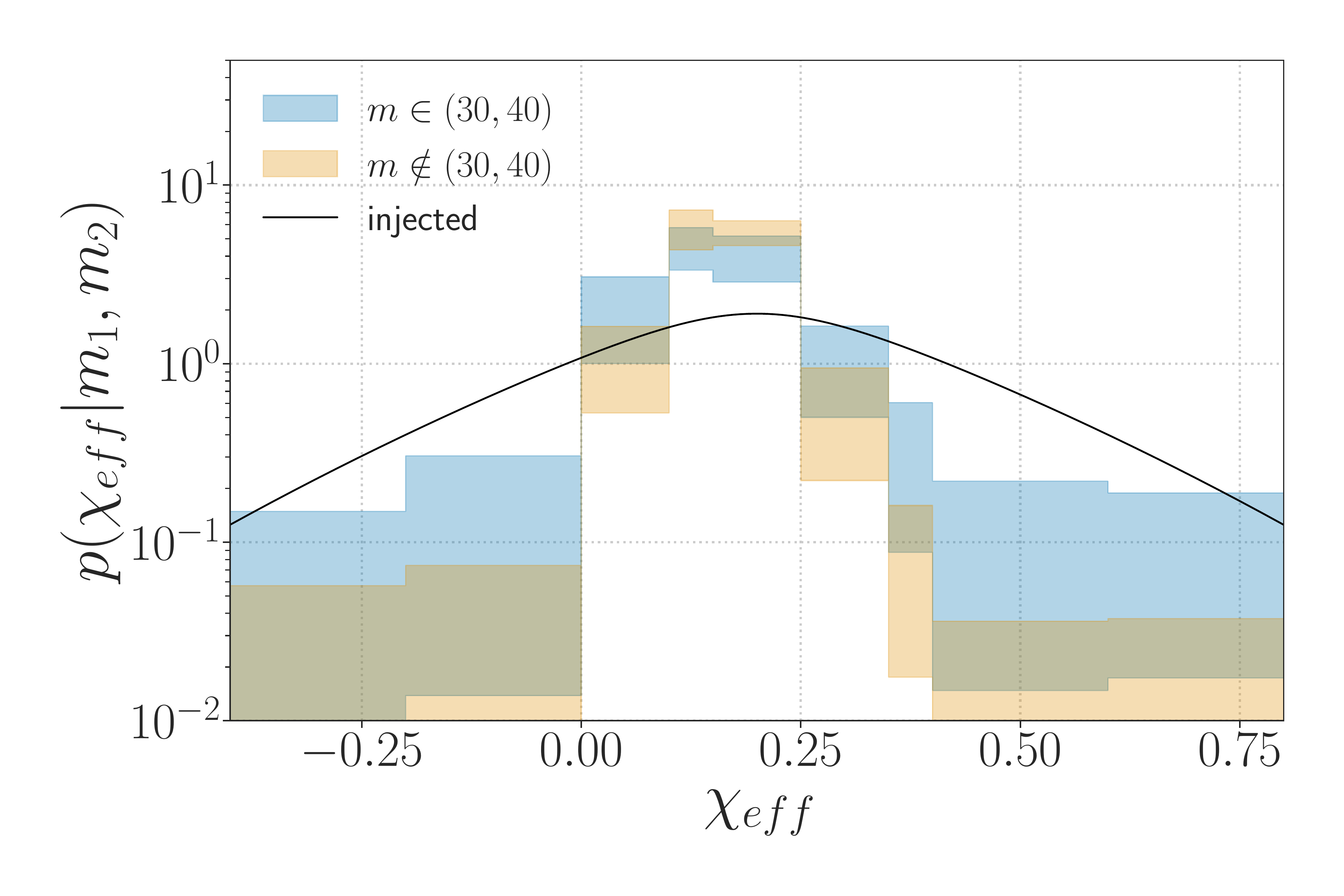}
\caption{\label{fig:sims}Conditional mass-spin distributions for the three simulated catalogs. The uncorrelated population is shown on the \textit{left}, the one with $\chi_{eff}\mbox{-}q$ correlations in the middle and the one with spin-redshift correlations on the right. Note that for the right panels, the reconstructed $\chi_{eff}$ distributions are not expected to match the injected population since our model assumes the joint $\chi_{eff}-z$ distribution to be separable, i.e., $p(\chi_{\text{eff}},z)=p(\chi_{\text{eff}})p(z)$.}
\end{center}
\end{figure*}
To validate the robustness of our results obtained from real data we test our model on large simulated catalogs. These datasets are generated by drawing BBH parameters from a known population distribution and injecting the corresponding signals into detector noise realizations. We use Bayesian parameter estimation on the injected signals using the \texttt{BILBY} package~\citep{Ashton:2018jfp,Romero-Shaw:2020owr}, its \texttt{DYNESTY}-based nested sampler~\citep{Speagle:2019ivv}, and the aligned spin waveform approximant known as IMPRphenomD~\citep{Husa:2015iqa,Khan:2015jqa}. We use LVK's noise power spectral densities that were measured during the first three months of its third observing run~\citep{ligoLIGOT2000012v2Noise} and draw 276 events for each simulated population which is roughly 4 times the size of the GWTC-3 BBH catalog. We use sensitivity injections found during the third observing run as publicly released by the LVK~\citep{gwosco3} to compute the detectable time-volumes required for analyzing these simulated populations.

\subsection{Testing for false positives}

We first simulate three BBH populations each of which is characterized by a \textsc{Powerlaw + Peak} model for the marginal primary mass-distribution~\citep{plpp0}, a mass-ratio dependent pairing function \citep{Fishbach:2019bbm}, a power-law in one plus redshift for the evolution of the merger rate~\citep{redshift_ev}, and a Gaussian distribution for effective inspiral spins. In the first population, we impose no intrinsic correlations between masses, spins, and redshifts. In the second one, we introduce an anti-correlation between mass-ratio and effective inspiral spins by making the mean of the $\chi_{\rm{eff}}$ Gaussian a linear function of mass-ratio with a negative slope~\citep{Callister:2021fpo}. In the third population we allow the $\chi_{\rm{eff}}$ distribution to broaden with redshift, similar to the models of \cite{Biscoveanu:2022qac}.

\begin{widetext} 
All three population distributions can be expressed in the following form:
\begin{equation}
    \frac{dN}{dm_1dqdzd\chi_{\rm{eff}}} =N_{0}\left(\lambda\frac{(1-\alpha)m_{1}^{-\alpha}}{m_{\rm{max}}^{1-\alpha}-m_{\rm{min}}^{1-\alpha}}+(1-\lambda)\frac{e^{-\frac{1}{2}\left(\frac{m_1-\mu_m}{\sigma_m}\right)^2}}{\sqrt{2\pi}\sigma_m}\right)\frac{(1+\beta)q^{\beta}}{1-\left(\frac{m_1}{m_{\rm{min}}}\right)^{\beta+1}} \frac{dV}{dz}(1+z)^{\kappa-1} \frac{e^{-\frac{1}{2}\left(\frac{\chi_{\rm{eff}}-\mu_{\chi}+\delta\mu_{\chi}(q-0.5)}{\sigma_{\chi}+\delta\sigma_{\chi}(z-0.001)}\right)^2}}{\sqrt{2\pi}(\sigma_{\chi}+\delta\sigma_{\chi}(z-0.001))}\label{eq:sims}
\end{equation}
where $q=\frac{m_2}{m_1}$ is the mass-ratio and  $N_0$ the total number of BBHs that have occured through out the observation time. The hyper-parameters $(m_{\rm{min}},m_{\rm{max}},\lambda,\alpha,\beta,\mu_m,\sigma_m,\kappa,\mu_{\chi},\delta\mu_{\chi},\sigma_{\chi},\delta\sigma_{\chi})$ corresponding to the three simulated populations are listed in Table~\ref{table:true-vals}.
\end{widetext}

The reasons for choosing these three populations to conduct our validation study are as follows. Using the un-correlated mock dataset, we demonstrate that the kind of mass-spin correlations we are seeing in real data are not spuriously manifesting from any artifacts built into our model, nor from the correlations between the measurement uncertainties of the BBH observables of interest~(left panels of Fig.~\ref{fig:sims}). Using the simulated population with intrinsic $q\mbox{-}\chi_{\rm{eff}}$ correlation, we demonstrate that such a correlation cannot manifest into multiple subpopulations in the marginal $m_1\mbox{-}\chi_{\rm{eff}}$ plane~(middle panels of Fig.~\ref{fig:sims}). Finally, using the population with intrinsic spin-redshift correlations we validate that our results for GWTC-3 are not indicative of spin-redshift correlations but instead of new astrophysics underlying BBH formation~(right panels of Fig.~\ref{fig:sims}). 

We note that for the simulated dataset with intrinsic spin-redshift correlation, the reconstructed $\chi_{\text{eff}}$ distributions are not expected to match the injected population since our model assumes the joint $\chi_{\text{eff}}\mbox{-}z$ distribution to be separable, i.e., $p(\chi_{\text{eff}},z)=p(\chi_{\text{eff}})p(z)$. The point of this validation study is to instead show that the ignored spin-redshift correlations cannot manifest into the kind of mass-spin correlations we are seeing, in the context of our binned model. On the other hand, for the simulations with intrinsic spin-mass ratio correlation, we choose a higher signal to noise ratio~(SNR) threshold than the other two to make sure that the poor measurability of both mass ratio and effective spin at low SNR does not bias the inferred shapes.

\begin{table}[h]
\centering
\begin{tabular}{cccc}
\hline
\hline
Parameter & Uncorrelated & Spin-mass-ratio & Spin-redshift \\
\hline
$m_{\rm{min}}/M_{\odot}$ & 6.0 & 6.0 & 6.0 \\
$m_{\rm{max}}/M_{\odot}$ & 70.0 & 70.0 & 70.0\\
$\lambda$ & 0.04 & 0.04 & 0.04\\
$\alpha$ & 4.0 & 4.0 & 4.0\\
$\beta$ & 1.5 & 1.5 & 1.5\\
$\mu_m/M_{\odot}$ & 35.0 & 35.0 & 35.0\\
$\sigma_m/M_{\odot}$ & 4.0 & 4.0 & 4.0\\
$\kappa$ & 3.0 & 3.0 & 3.0\\
$\mu_{\chi}$ & 0.2 & 0.2 & 0.2\\
$\delta\mu_{\chi}$ & 0.0 & $-0.5$ & 0.0\\
$\sigma_{\chi}$ & $10^{-1.8}$ & $10^{-1.8}$ & $10^{-0.85}$\\
$\delta\sigma_{\chi}$ & 0.0 & 0.0 & $10^{-0.88}$\\
\hline
\end{tabular}
\caption{True values for the hyper-parameters characterizing the underlying population in Eq.~\eqref{eq:sims} for the three simulated catalogs. The parameters controlling the correlations are chosen to have best-fit values from \cite{KAGRA:2021duu} }
\label{table:true-vals}
\end{table}

\subsection{Recovering different kinds of mass-spin correlations}
\begin{figure*}[htt]
\begin{center}
\includegraphics[width=0.32\textwidth]{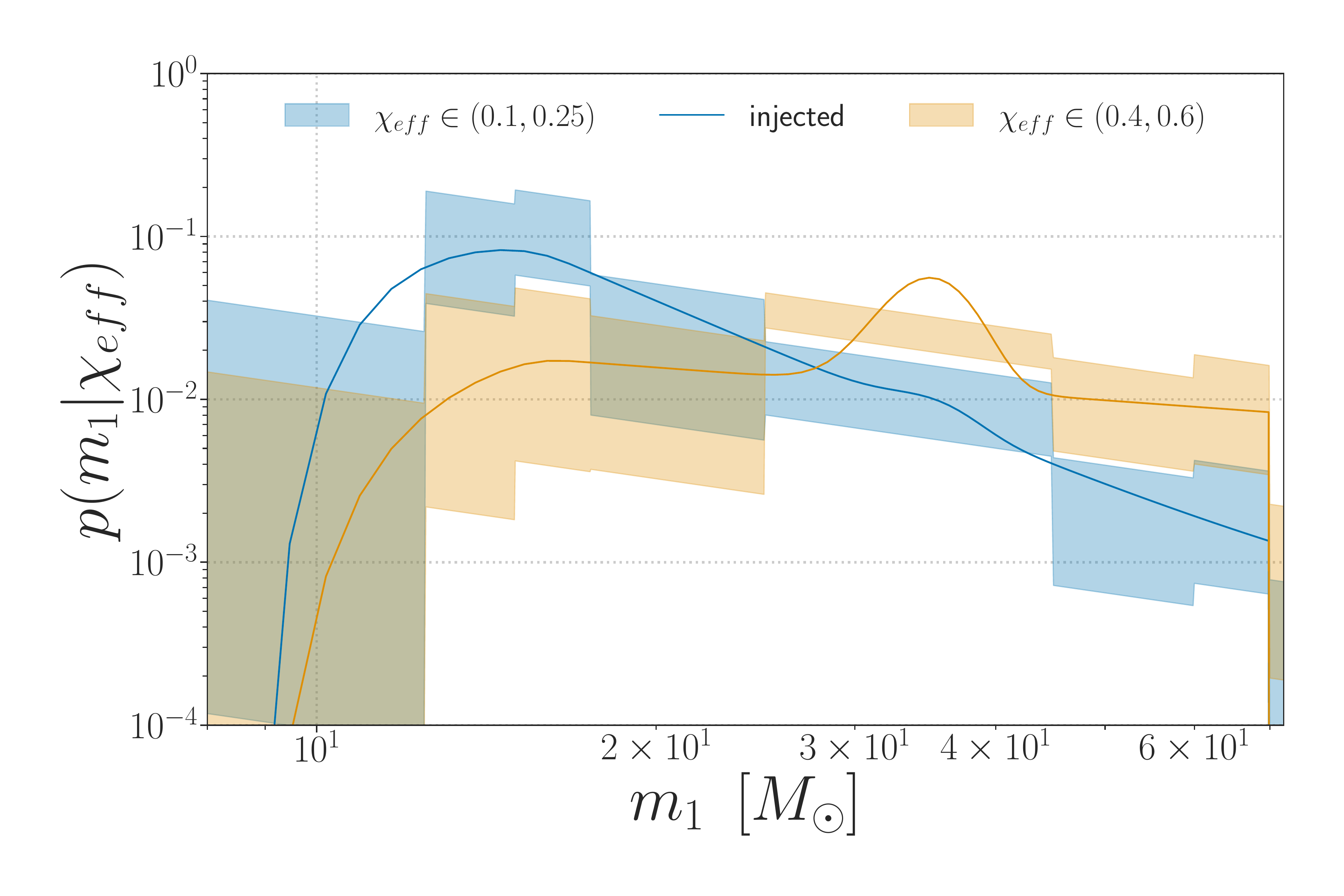}
\includegraphics[width=0.32\textwidth]{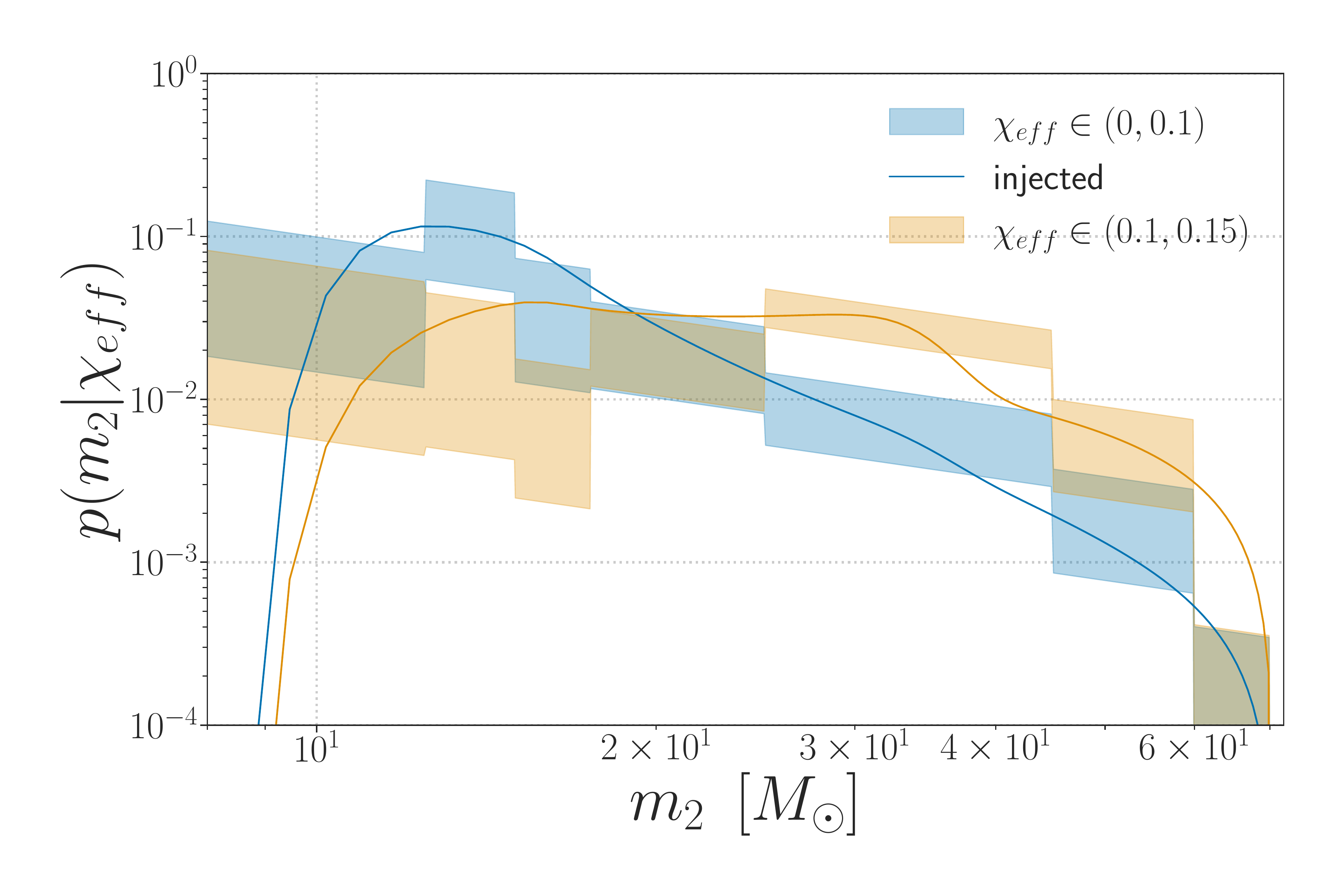}
\includegraphics[width=0.32\textwidth]{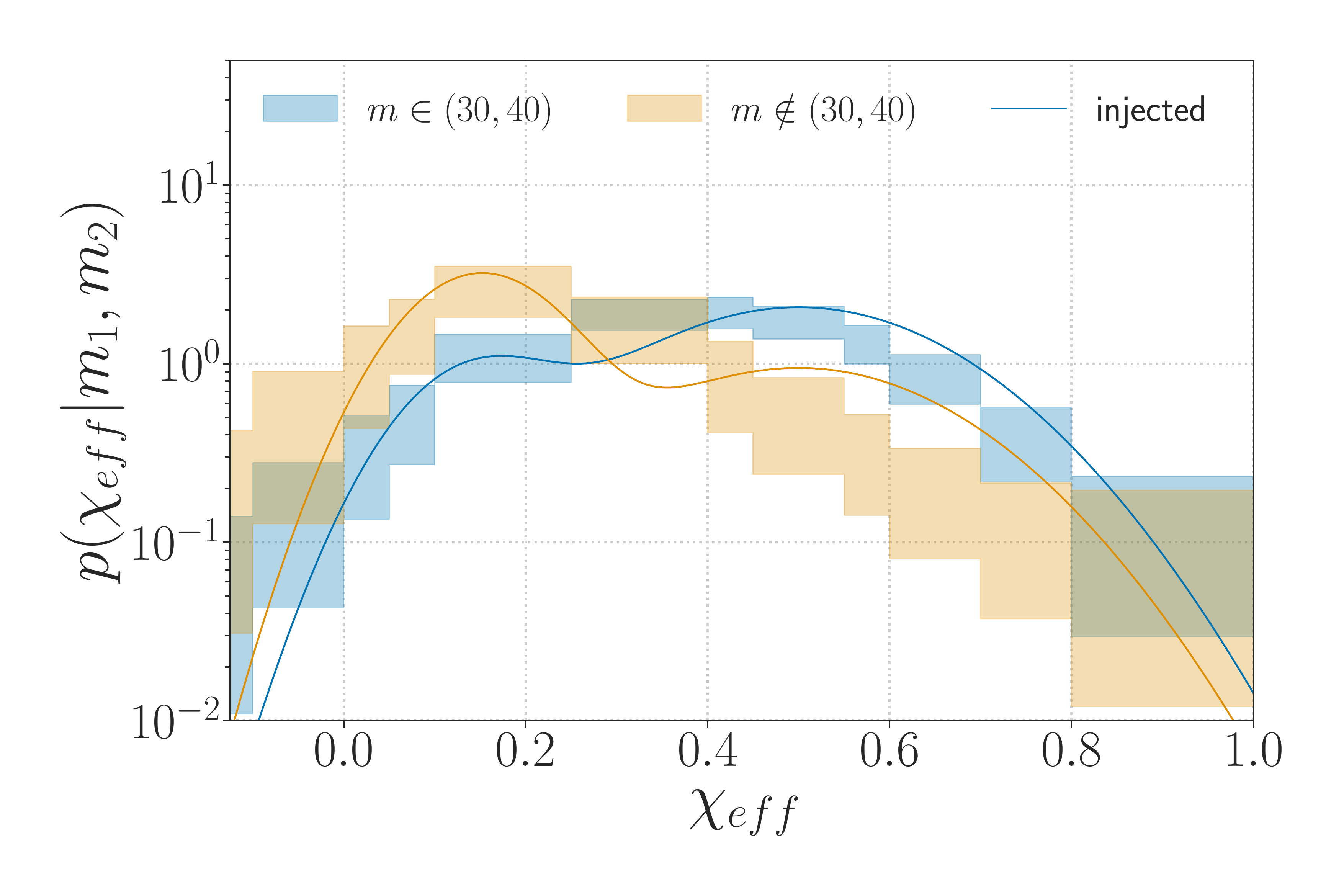}
\includegraphics[width=0.32\textwidth]{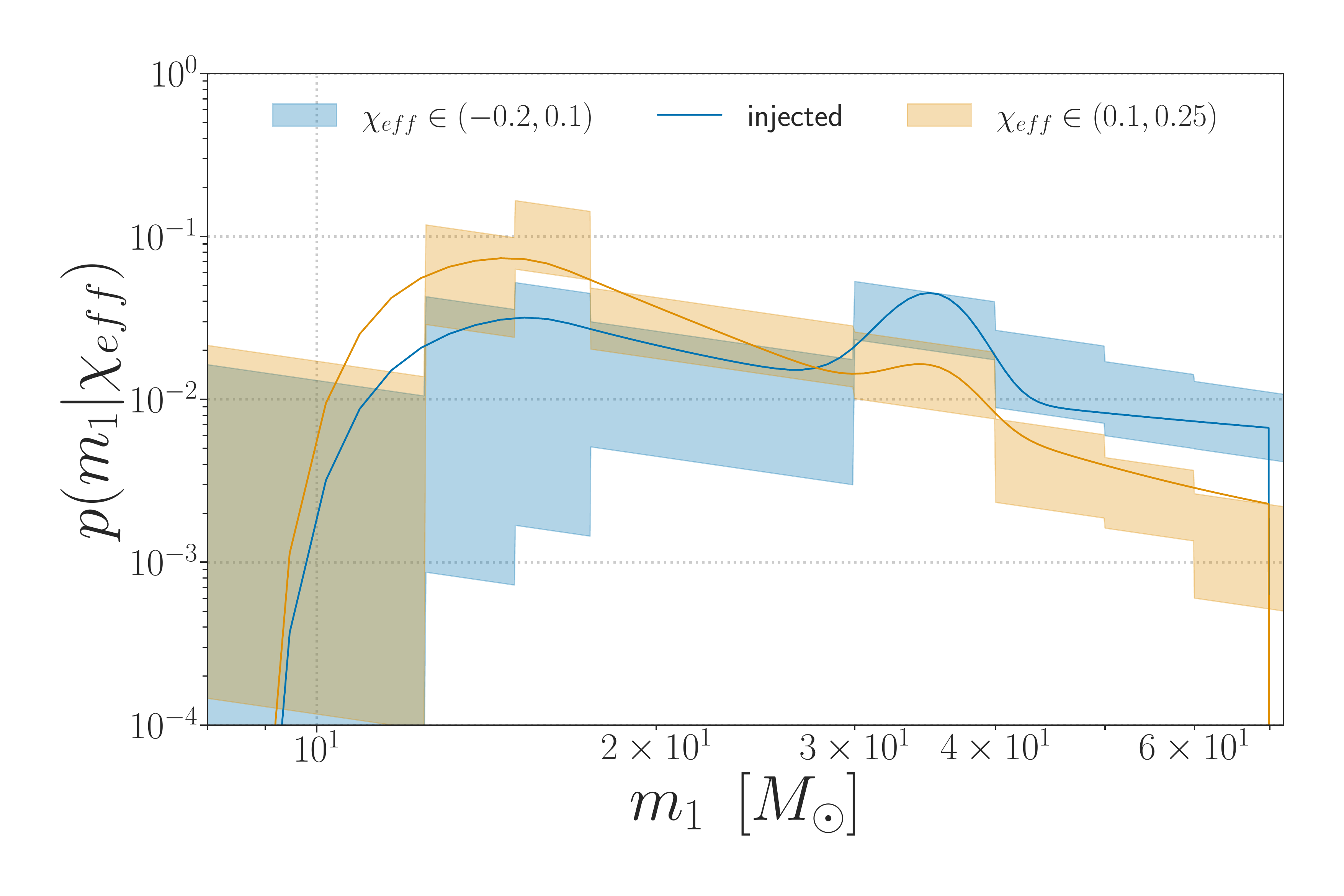}
\includegraphics[width=0.32\textwidth]{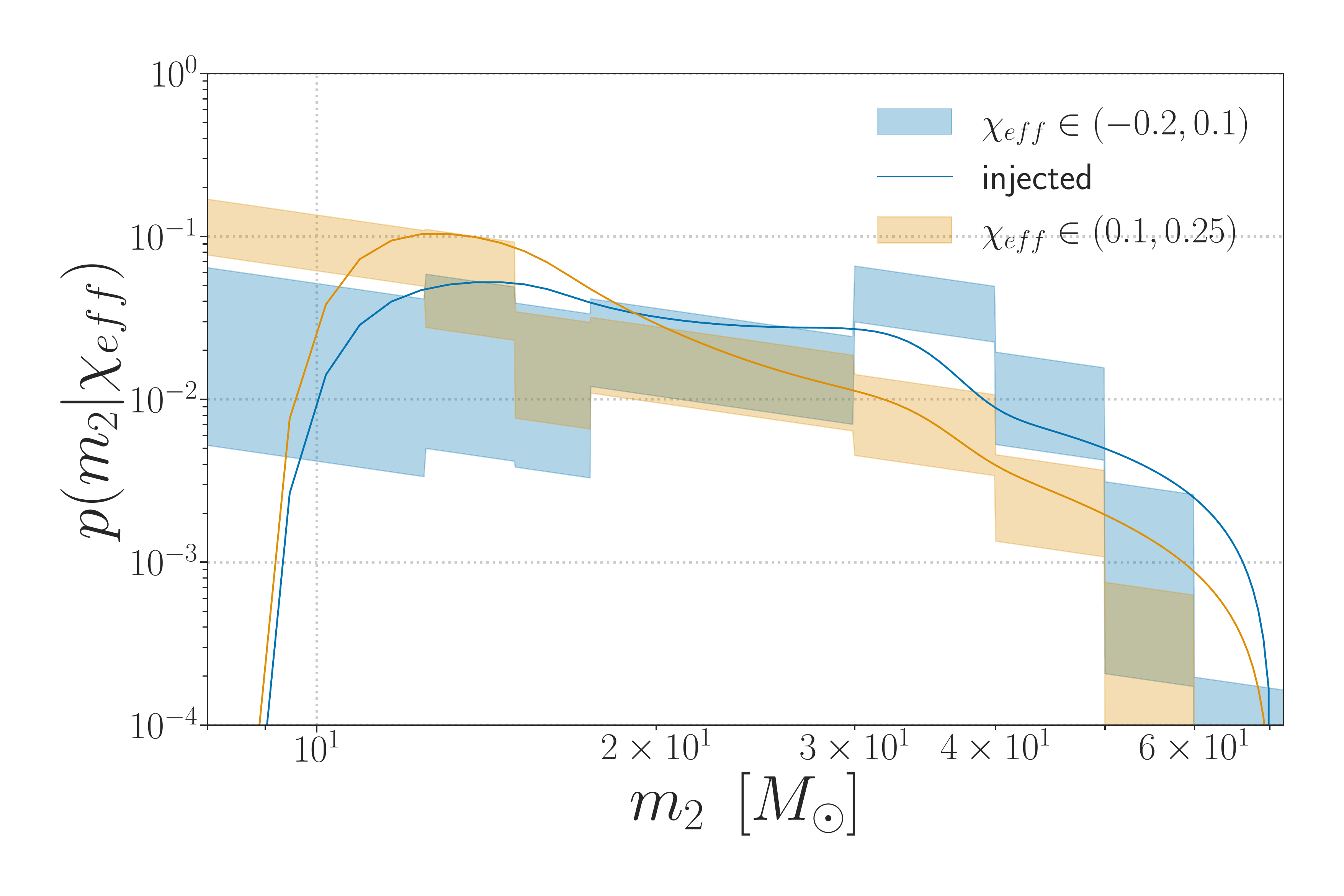}
\includegraphics[width=0.32\textwidth]{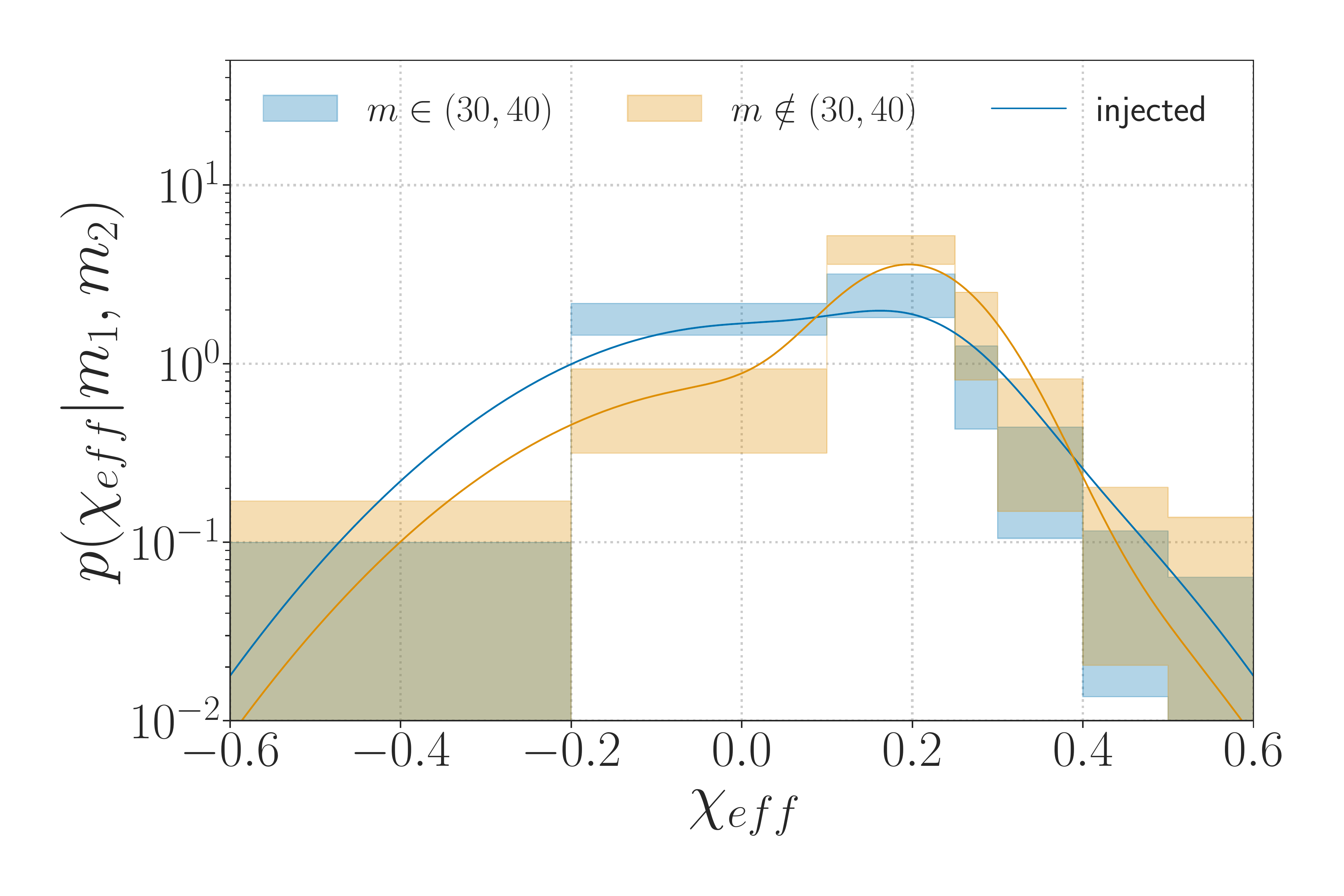}
\caption{\label{fig:sims2}Inferred conditional distributions for the two simulated catalogs with intrinsic mass-spin correlations resulting from the mixture of two subpopulations. The top and bottom pannel represrents the peak+aligned and peak+isotropic populations respectively from table~\ref{table:true-vals2} }
\end{center}
\end{figure*}

In the previous section, we had shown that various forms of $\chi_{\text{eff}}$ correlations which have already been found in GWTC-3, cannot manifest into the trends we have found in the same dataset. In this section we simulate two additional populations with intrinsic mass-spin correlations, similar to the trends we have found in GWTC-3, and show that our model can accurately recover injected conditional distributions of mass given effective spin and vice-versa. For the primary mass distribution, we choose a mixture of two subpopulations, one of which is a power-law without any peak and the other a Powerlaw+Peak with a very shallow spectral index and a large peak fraction. Each of these subpopulations is associated with a Gaussian $\chi_{\text{eff}}$ distribution whose mean and standard deviations are different from that of the other subpopulations. The mass-ratio distribution and redshift evolution of the merger rate are both chosen to be the same as that of the mock populations in the previous section.
\begin{widetext}
These two population distributions can be expressed in the following form:
    \begin{equation}
        \frac{dN}{dm_1dqdzd\chi_{\rm{eff}}} =\sum_{i=1}^{2}f_iN_{0,i}\left(\lambda_i\frac{(1-\alpha)m_{1}^{-\alpha_i}}{m_{\rm{max}}^{1-\alpha_i}-m_{\rm{min}}^{1-\alpha_1}}+(1-\lambda_i)\frac{e^{-\frac{1}{2}\left(\frac{m_1-\mu_{m,i}}{\sigma_{m,i}}\right)^2}}{\sqrt{2\pi}\sigma_{m,i}}\right)\frac{(1+\beta)q^{\beta}}{1-\left(\frac{m_1}{m_{\rm{min}}}\right)^{\beta+1}} \frac{dV}{dz}(1+z)^{\kappa-1} \frac{e^{-\frac{1}{2}\left(\frac{\chi_{\rm{eff}}-\mu_{\chi,i}}{\sigma_{\chi,i}}\right)^2}}{\sqrt{2\pi}(\sigma_{\chi,i})}\label{eq:sims2}
    \end{equation}
    where the hyper-parameters corresponding to the two simulated populations are listed in Table~\ref{table:true-vals2}
\end{widetext}
We set the mean of the effective-spin Gaussian associated with the non-zero peak mass distribution to be 0.5 in one of the simulated datasets~(peak+aligned) and 0 in the other~(peak+isotropic, table~\ref{table:true-vals2}). The subpopulation without the peak is chosen to have an effective spin distribution with mean 0.2 for both cases. We analyze both mock catalogs with our model and show that it can accurately reconstruct the injected distribution~(Figure~\ref{fig:sims2}). In other words, for GWTC-3, had the subpopulation with the $30-40 M_{\odot}$ feature been associated with more positively aligned spin distributions, our model would have recovered it correctly.

\begin{table}[h]
\centering
\begin{tabular}{ccc}
\hline
\hline
Parameter & Peak+aligned & Peak+isotropic \\
\hline
$m_{\rm{min}}/M_{\odot}$ & 8.0 & 8.0 \\
$m_{\rm{max}}/M_{\odot}$ & 70.0 & 70.0 \\
$\lambda_1$ & 0 & 0 \\
$\lambda_2$ & 0.4 & 0.4 \\
$\alpha_1$ & 3 & 3\\
$\alpha_2$ & 0.5 & 0.5\\
$\beta$ & 1.5 & 1.5 \\
$\mu_m/M_{\odot}$ & 35.0 & 35.0 \\
$\sigma_m/M_{\odot}$ & 3.0 & 3.0 \\
$\kappa$ & 3.0 & 3.0 \\
$\mu_{\chi,1}$ & 0.2 & 0.2\\
$\mu_{\chi,2}$ & 0.5 & 0\\
$\sigma_{\chi,1}$ & $10^{-1.1}$ & $10^{-1.1}$  \\
$\sigma_{\chi,2}$ & $10^{-0.7}$ & $10^{-0.8}$ \\
$f{1}$ & 0.6 & 0.6 \\
$f{2}$ & 0.4 & 0.4 \\
\hline
\end{tabular}
\caption{True values for the hyper-parameters characterizing the underlying population in Eq.~\eqref{eq:sims2} for the three simulated catalogs. The parameters controlling the correlations are chosen to have best-fit values from \cite{KAGRA:2021duu} }
\label{table:true-vals2}
\end{table}

\section{Astrophysical Implications}
\label{sec:astro}
We have found evidence for a subpopulation of BBH mergers in the mass range $30-40 \ M_{\odot}$ with preference for a symmetric $\chi_{\rm{eff}}$ population distribution. As explored by different studies, the $30-40 \ M_{\odot}$ feature seen in the GWTC-3 BBH inferred mass spectrum with models such as \textsc{Powerlaw + Peak}, cannot be explained by the PISN/PPISN process as this feature is expected to be above the $50 M_{\odot}$ mass range \citep{Hendriks:2023yrw,PPISN,Thrane:2018qnx,Golomb:2023vxm}.  In the rest of this section we consider different proposed formation channels which could be responsible for this subpopulation.

%KB hacked this section

Isolated binary evolution formation channels generally rely on Roche-overflow mass transfer or close initial binary separations to produce BBHs that are close enough to merge within a Hubble time. In the case of Roche overflow, the mass transfer can proceed stably in all cases potentially leading to mass ratio reversals \citep[e.g.][]{2021ApJ...922..110G,Broekgaarden:2022nst,2022ApJ...940..184V,2023NatAs...7.1090B} or unstably leading to common envelope evolution \citep[e.g.][]{Mapelli:2018uds,Zevin:2020gbd,Wong:2020ise,2022ApJ...935..126B,vanSon:2022ylf}. In the case of close initial binary separations, the stellar components are expected to rotate with periods synchronized to the orbital period such that the stars evolve in a chemically homogeneous fashion and thus do not experience significant radius increases over their lifetimes \citep[e.g.][]{Mandel:2015qlu,Marchant:2016wow,2016MNRAS.460.3545D}. While common envelope evolution is primarily expected to produce BBHs with masses $<30 M_{\odot}$, both stable mass transfer and chemically homogeneous evolution could produce BBHS in this mass range. However, due to the coevolution of the stars in the binary, our results suggest that most of the merger rate contribution from either of these channels are likely not significantly contributing to the $30-40 \ M_{\odot}$ merger rate due to the preference for a symmetric $\chi_{\rm{eff}}$ population distribution \citep[e.g,][]{2020A&A...635A..97B,2023MNRAS.520.5724B}.

Dynamically formed BBHs can originate from a wide variety of astrophysical environments from triple star systems \citep[e.g.,][]{2017ApJ...841...77A,2022ApJ...937...78M} to young stellar clusters \citep[e.g.,][]{2019MNRAS.487.2947D,2020MNRAS.498..495D}, globular clusters \citep[e.g.,][]{Rodriguez:2018pss,2022MNRAS.511.5797M}, and nuclear star clusters \citep[e.g.][]{Antonini:2012ad, 2017ApJ...846..146P}, as well as the disks of active galactic nuclei (AGN) \citep[e.g.][]{Bartos:2016dgn,2017MNRAS.464..946S,2020MNRAS.494.1203M}. While each of these environments are capable of producing masses near $30-40 \ M_{\odot}$, only globular clusters are predicted to produce BBHs with a mass distributions that peaks near $30-40 \ M_{\odot}$ \emph{and} a $\chi_{\rm{eff}}$ distribution that is symmetric about 0. In the case of nuclear star clusters the mass peaks toward lower values, while in the case of young stellar clusters the mass peaks toward higher values \citep{Cheng:2023ddt,2022MNRAS.511.5797M}. In the case of BBH formation in AGN disks, the mass and spin distributions are heavily influenced by the existence (or lack) of migration traps which predominantly produce low-mass-ratio mergers \citep[e.g.,][]{2016ApJ...819L..17B, 2020MNRAS.494.1203M}. 

%Isolated binary formation channels rely on generally two mechanisms to produce BBH binaries that can merger within a Hubble time. Other than common envelope evolution which is expected to yield BHs with masses $<30 M_{\odot}$, field binaries can form merging BBHs by undergoing 1) stable mass transfer, which tends to form BHs at higher masses and could potentially explain the $30-40 \ M_{\odot}$ mass feature \citep{Mandel:2015qlu,Broekgaarden:2022nst,vanSon:2022ylf}, and 2) chemically homogeneous evolution, which could also explain the peak at $30-40 \ M_{\odot}$\citep{Cheng:2023ddt}. However, our results suggest that most of the merger rate contribution from either of these field binary formation channels are likely not contributing significantly to $30-40 \ M_{\odot}$ merger rate due to the preference for a symmetric $\chi_{\rm{eff}}$ population distribution, where as, these two mechanisms lead to spinning binaries with aligned spins, leading to a $\chi_{\rm{eff}}>0$ population preferentially \citep{Cheng:2023ddt,Rodriguez:2018pss,Rodriguez:2021qhl,Gerosa:2021mno}.  Therefore, if we assume that the BBH population distribution from GWTC-3 is composed predominantly of two formation channels, e.g. field binaries and dynamically assembled binaries \citep{Zevin:2020gbd}. We have suggestive evidence that the $30-40 \ M_{\odot}$ is likely associated to BBH mergers that originated in dense stellar environments.

%\textcolor{orange}{perhaps mention that STM can produce 30-40 but they will be alisgned and non-negligibly spinnins, cite(Briel et al 2023, Zevin and Bavera 2022?)}

We can isolate the $30-40 \ M_{\odot}$ subpopulation and compute the measured local merger rate $R_0^{\rm{peak}}$ (at $z=0$) under various assumptions. First, we compute the contribution by marginalizing over the remaining mass bins and across the entire $\chi_{\rm{eff}}$ range. Second, we can compute this contribution $R_0^{\rm{low }\mathrm{\chi}}$ over the range $0.05<\chi_{\rm{eff}}<0.1$ only and also provide an estimate for the local merger rate $R_0^{\rm{other}}$ in the complementary mass and $\chi_{\rm{eff}}$ bins, a proxy, for the merger rate due to field binaries and other formation channels which we neglect. In the first case we find,  $R_0^{peak} = 2.04_{-0.73}^{+0.79} \ \rm{Gpc^{-3}yr^{-1}}$ with the merger rate of other binaries to be  $R_0^{\rm{other}} = 22.4_{-9.4}^{+10.2} \ \rm{Gpc^{-3}yr^{-1}}$. Similarly, for the second more restrictive case we consider, we find, $R_0^{\rm{low }\mathrm{\chi}}  = 1.09^{+0.81}_{-0.95} \ \rm{Gpc^{-3}yr^{-1}}$, while for the remaining binaries we find, $R_0^{\rm{other}} = 23.27^{+10.51}_{-9.50}$. We measure a total merger rate of $R_0 = 24.4_{-9.2}^{+10.7} \ \rm{Gpc^{-3}yr^{-1}}$ for reference. From the estimated local merger rates we estimate the percentage of binaries contributing to the $30-40 \ M_{\odot}$ feature to be in the range of $1-8\%$.

Under the assumption that the $30-40 \ M_{\odot}$ and $0.05<\chi_{\rm{eff}}<0.1$ regions contain dominant contributions from the dynamical channel, our results indicate that we find merger rates in the range $R_0^{\rm{dyn}} = 0.1-2.8 \ \rm{Gpc^{-3}yr^{-1}}$ for these subpopulations. We also find merger rates in the range $R_0^{\rm{other}} = 13-34 \ \rm{Gpc^{-3}yr^{-1}}$ for the remaining contribution due to other formation channels. If we assume that this subpopulation is due entirely to mergers occuring in GCs, we expect to have merger rates in the range $1-4 \ \rm{Gpc^{-3}yr^{-1}}$ and hence find consistency with theoretical models \citep{Mapelli:2021gyv,Rodriguez:2018pss}. However, we find a preference towards the lower and more pessimistic end of the rate predictions. We note that our measured rates are also broadly consistent with other dynamical formation environments such as young star clusters and nuclear star clusters. However, the mass feature being at $30-40 \ M_{\odot}$, makes GCs most consistent with our findings \citep{Mapelli:2021gyv}.

%Another possibility is the AGN formation channel, current models note that negative chieff are disfavoured due to. However, although rare, it can happen. Anything can happen in AGNs. Deal with AGNs later.

\section{Discussion and Conclusion}
\label{sec:conclusion}
%\textcolor{orange}{AR can take first shot once all results are finalized}
In this work, for the first time, we have characterized the joint distribution of BBH component masses and effective inspiral spins using a highly data-driven BGP-based population model. Using GWTC-3 data, we have found hints of multiple sub-populations in the BBH mass spectrum that are associated with different ranges of effective inspiral spins. The high (more positive) spin sub-population is found to have a powerlaw-like shape with no feature at the $30-40M_{\odot}$ range, which is consistent with the isolated binary formation channel and with the predictions of stellar evolution models that place the PPISN feature at $50M_{\odot}$ or higher~\citep{Hendriks:2023yrw}. The low (closer to zero) spin sub-population demonstrates a feature at the $30-40M_{\odot}$ range and is consistent with the expectations for BBHs formed dynamically in globular clusters, which are thought to peak around $30M_{\odot}$. These implications are corroborated by the conditional effective spin distributions that we have inferred. We have found that BBHs in the $30-40M_{\odot}$ range have an effecetive spin distribution more symmetric about zero as compared to that of other BBHs which correspond to a more positively skewed effective spin distribution. We have also computed the combined merger rates in the $30-40M_{\odot}$ and $0.05<\chi_{\rm{eff}}<0.1$ ranges and have found that they are consistent with the theoretical predictions for dynamically formed BBHs in globular clusters.

Our conclusions align broadly with the findings of \cite{Godfrey:2023oxb} who employ flexible mixture models to search for BBH subpopulations in GWTC-3 data, except a few subtle distinctions. We both identify two subpopulations in the mass spectrum one of which peaks near $10M_{\odot}$ and falls off steeply afterward, and the other one with a feature near the $30-40M_{\odot}$ range. However, \cite{Godfrey:2023oxb} find that the spin distributions of these two subpopulations are mostly consistent with each other, particularly the effective-spin distributions which they find to be almost completely overlapping. On the other hand, we find that the subpopulation that is comprised mostly of BBHs with at least one component in the $30-40M_{\odot}$ range has a significantly different effective spin distribution from the complementary subpopulation of binaries and hence arrive at slightly different astrophysical interpretations.

We have verified that our results are not susceptible to the inflexibilities that remain in our highly data-driven model. We have shown that our findings from GWTC-3 are robust against variations of binning choices as well as that of the redshift evolution parameter $\kappa$. Using large simulated catalogs corresponding to three different populations of BBHs, we have shown that the mass-spin correlations we are seeing cannot manifest spuriously due to artifacts built into our model assumptions, nor from any other kinds of correlations that are known to exist in the astrophysical BBH population and yet have been ignored in our simplified model. Additionally, using two simulated populations of BBHs with intrinsic mass-spin correlations, we have shown that if the subpopulation contributing dominantly to the $30-40M_{\odot}$ range had instead been associated with a more positively skewed effective spin distribution, our model would have accurately identified that trend.

At the end of LVK's ongoing fourth observing run, we expect our model to yield tighter constraints on the nature and existence of subpopulations in the BBH mass-spin distributions. As the size of GW catalogs continue to grow, we aim to generalize our model by constructing higher-dimensional BGPs that will self-consistently inform on mass-spin, spin-redshift and mass-redshift correlations in the BBH population and thereby constrain the underlying astrophysics of several BBH formation channels, beyond the limitations of parametric population modeling.

\section{Acknowledgements}
The authors would like to thank Thomas Callister, Vicky Kalogera, and Soumendra Roy for useful discussions and suggestions. This work was supported by the National Science Foundation award PHY-2207728. IMH is supported by a McWilliams postdoctoral fellowship at Carnegie Mellon University. The authors are grateful for computational resources provided by the LIGO Laboratory and supported by National Science Foundation Grants PHY-0757058 and PHY-0823459, PHY-0823459, PHY-1626190, and PHY-1700765. This material is based upon work supported by NSF's LIGO Laboratory which is a major facility fully funded by the National Science Foundation

\appendix
\section{Variation of the redshift evolution parameter}
\label{sec:kappa}
In this appendix we demonstrate that our results for GWTC-3 are robust against variations in the redshift evolution parameter $\kappa$, which we had originally fixed to be the median value found by \cite{KAGRA:2021duu} using the \textsc{Powerlaw+Peak} model from the same dataset $(\kappa=2.9)$.  Here we compare our results for the the joint mass-spin distributions for two additional choices of $\kappa$ which are taken to be the 90\% lower~$(\kappa=1.1)$ and upper~$(\kappa=4.6)$ bounds on it as reported by \cite{KAGRA:2021duu}.

\begin{figure*}[htt]
\begin{center}
\includegraphics[width=0.48\textwidth]{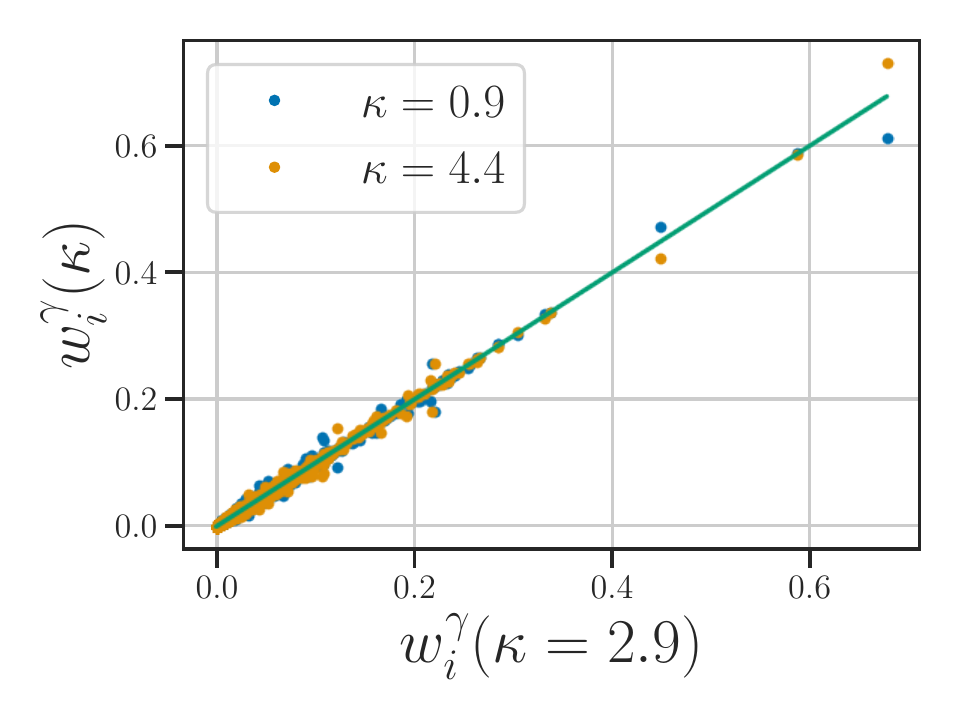}
\includegraphics[width=0.48\textwidth]{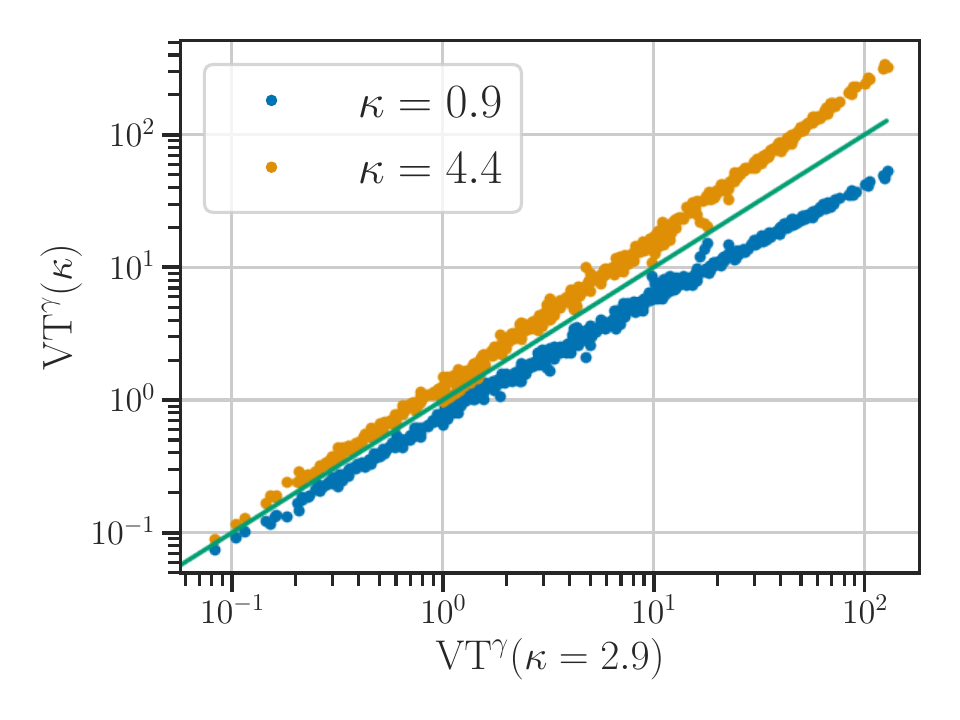}
\caption{\label{fig:weights-vts-vs-kappa}Variation of posterior weights~($w^{\gamma}_i$, \textit{left}) and detectable time-volumes~($\rm{VT}^{\gamma}_i$,\textit{right}) of Eq.~\eqref{likelihood} in each bin with $\kappa$.}
\end{center}
\end{figure*}

\begin{figure*}[htt]
\begin{center}
\includegraphics[width=0.32\textwidth]{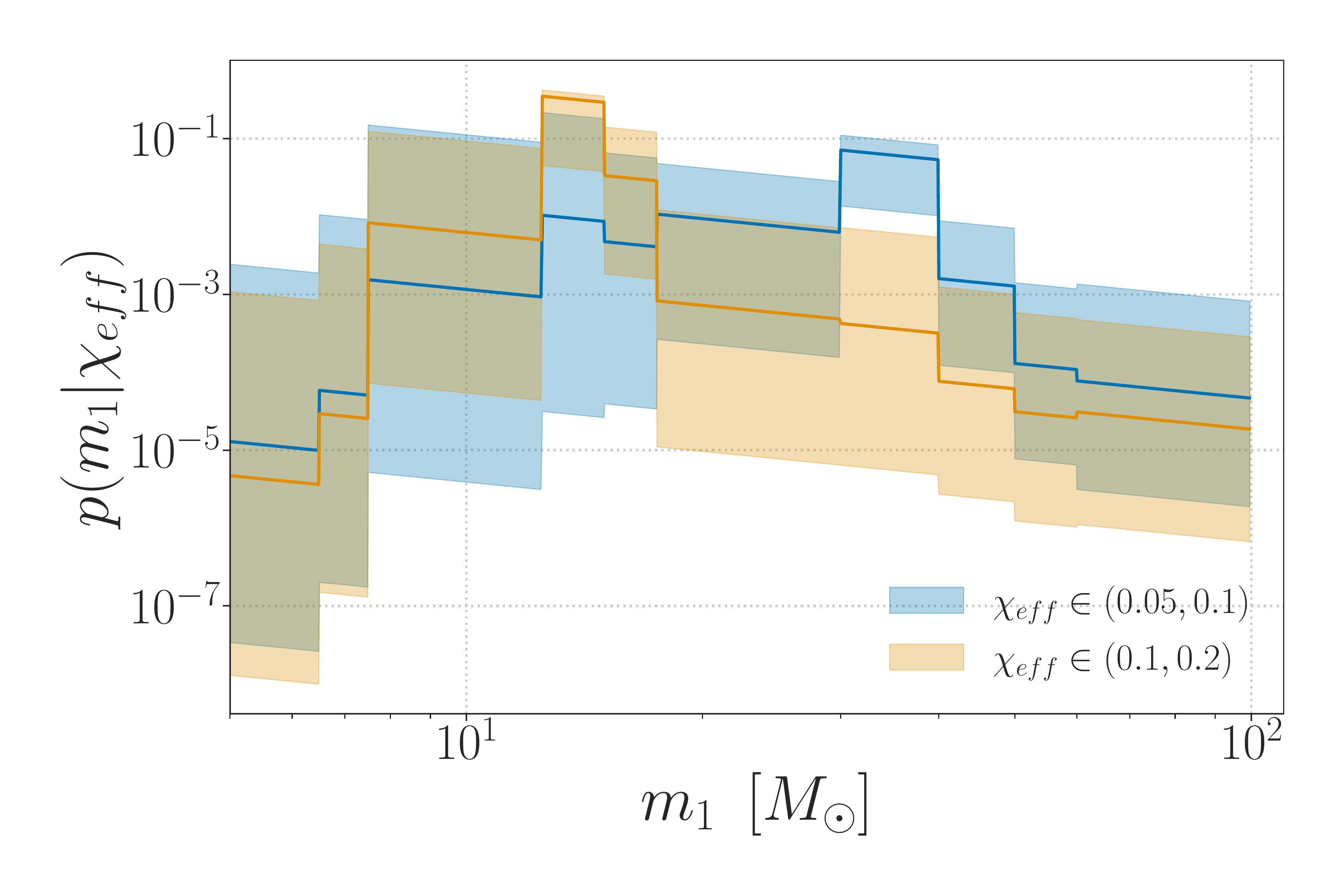}
\includegraphics[width=0.32\textwidth]{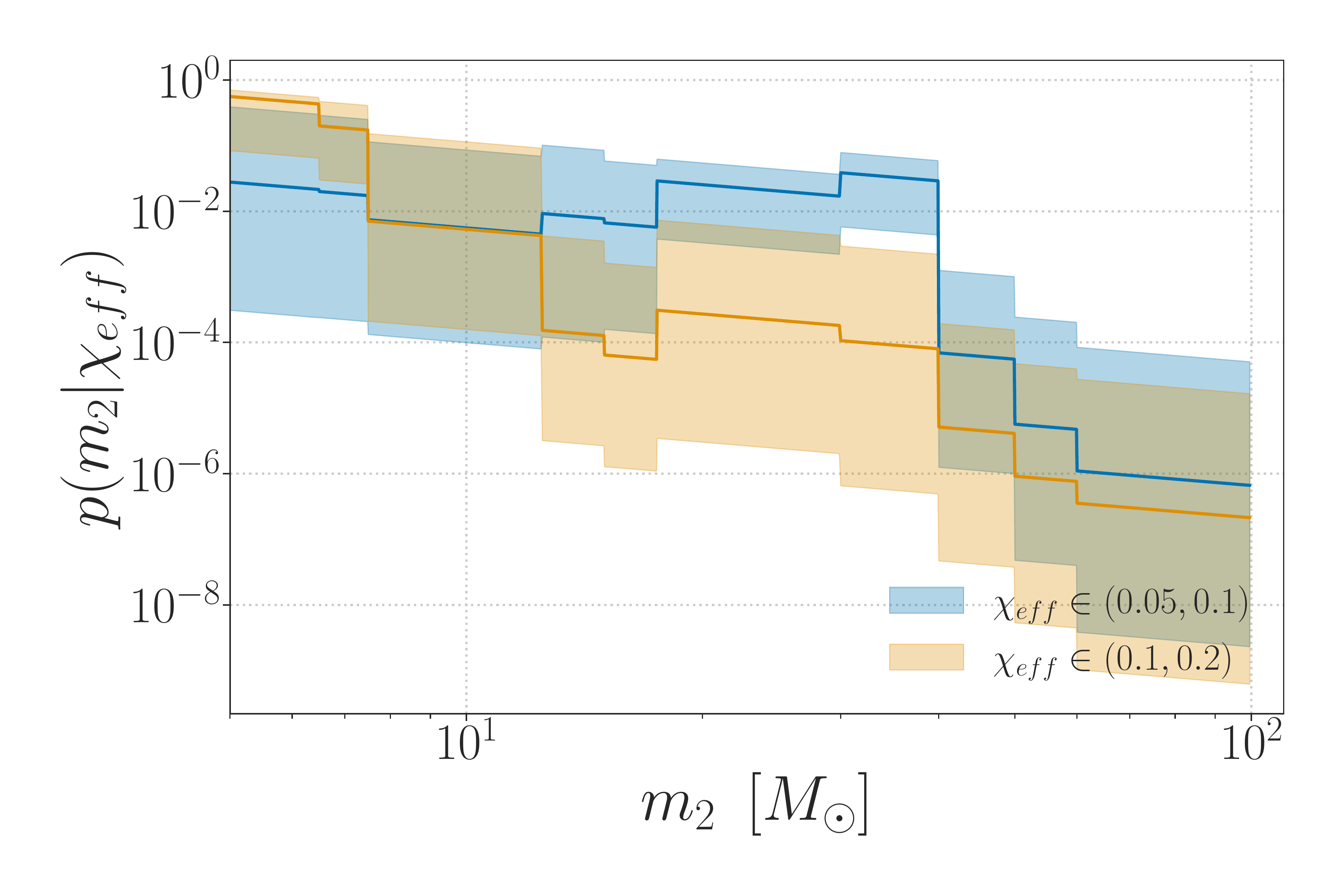}
\includegraphics[width=0.32\textwidth]{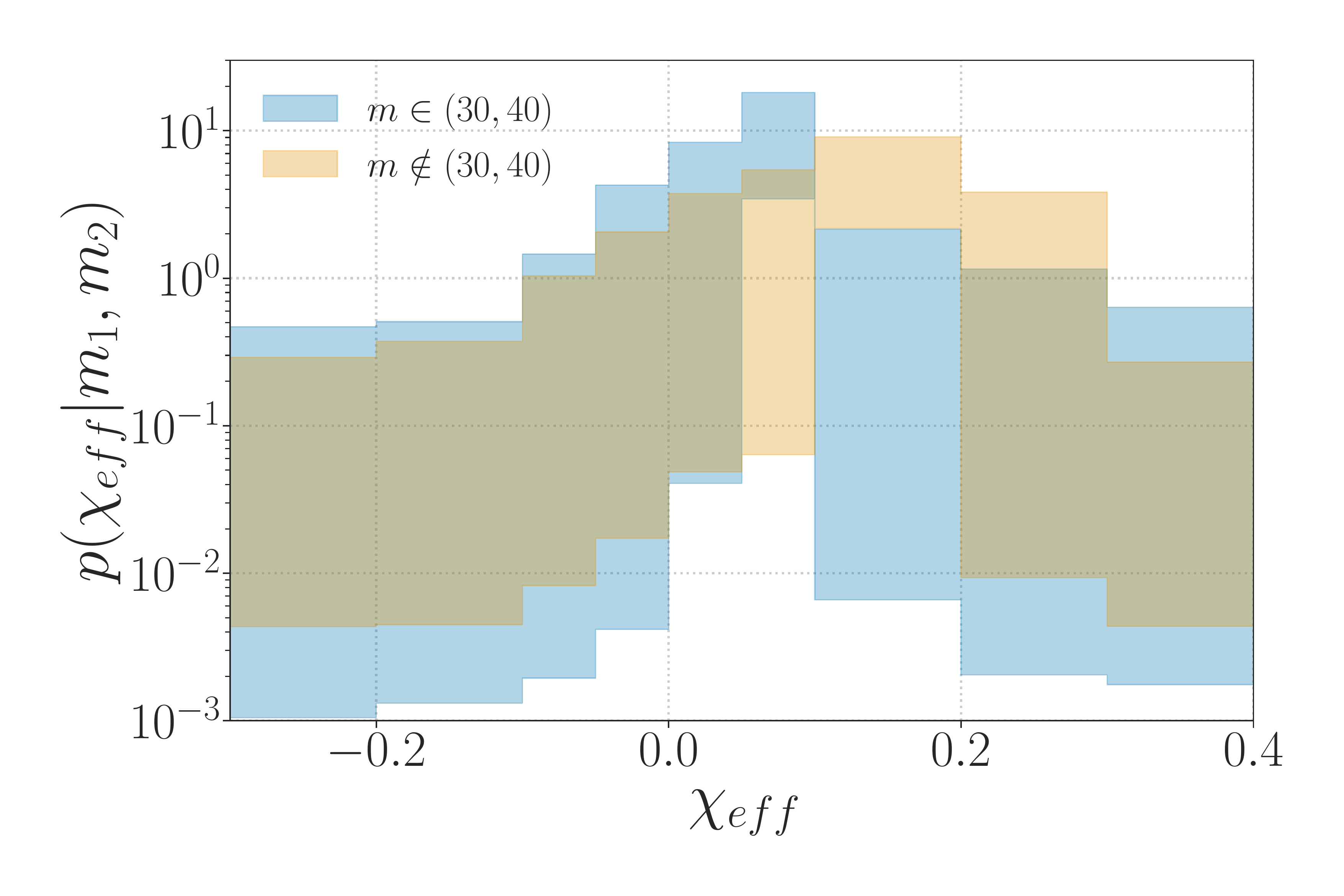}
\caption{\label{fig:kappa09} GWTC-3 results for $\kappa=1.1$}
\end{center}
\end{figure*}

\begin{figure*}[htt]
\begin{center}
\includegraphics[width=0.32\textwidth]{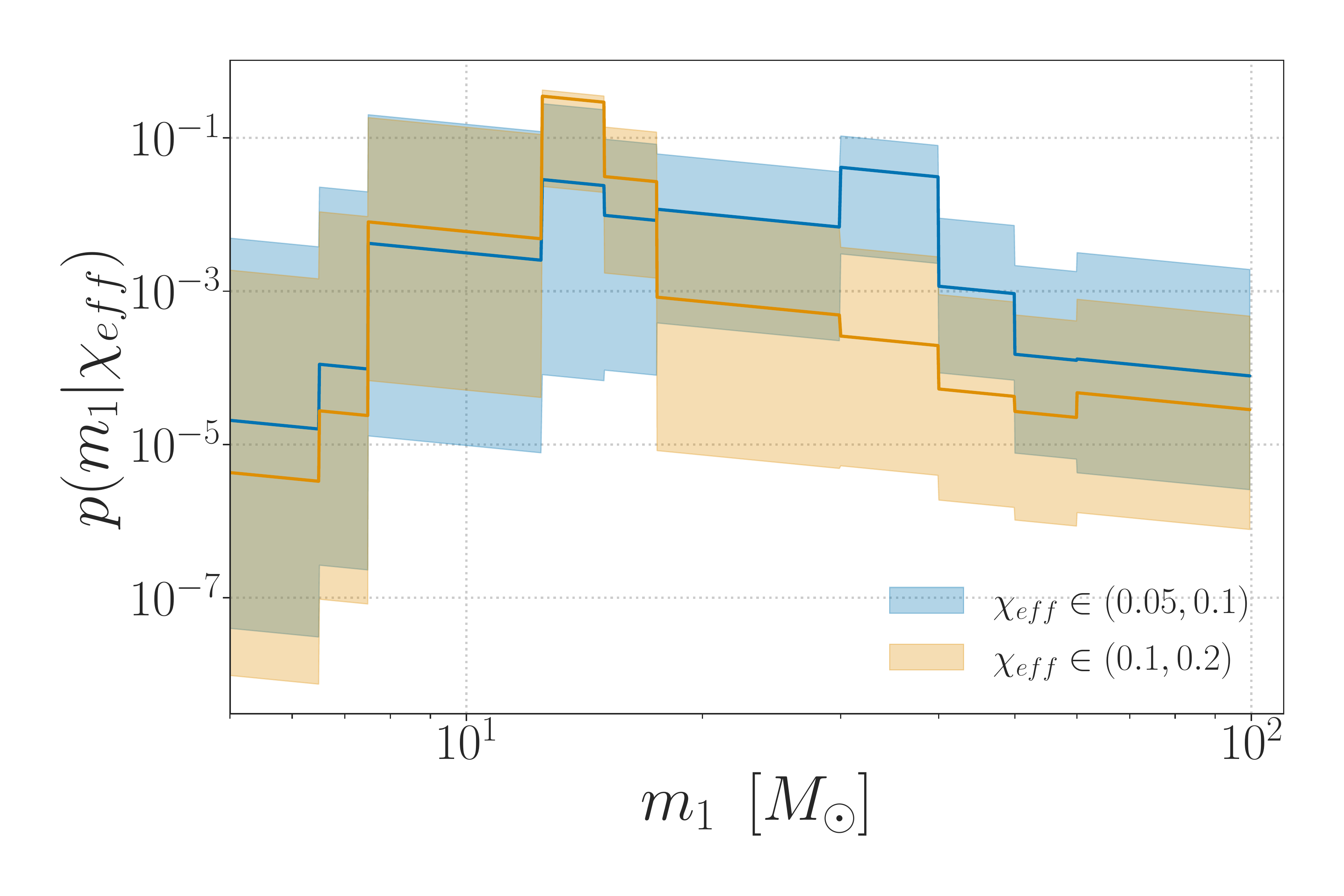}
\includegraphics[width=0.32\textwidth]{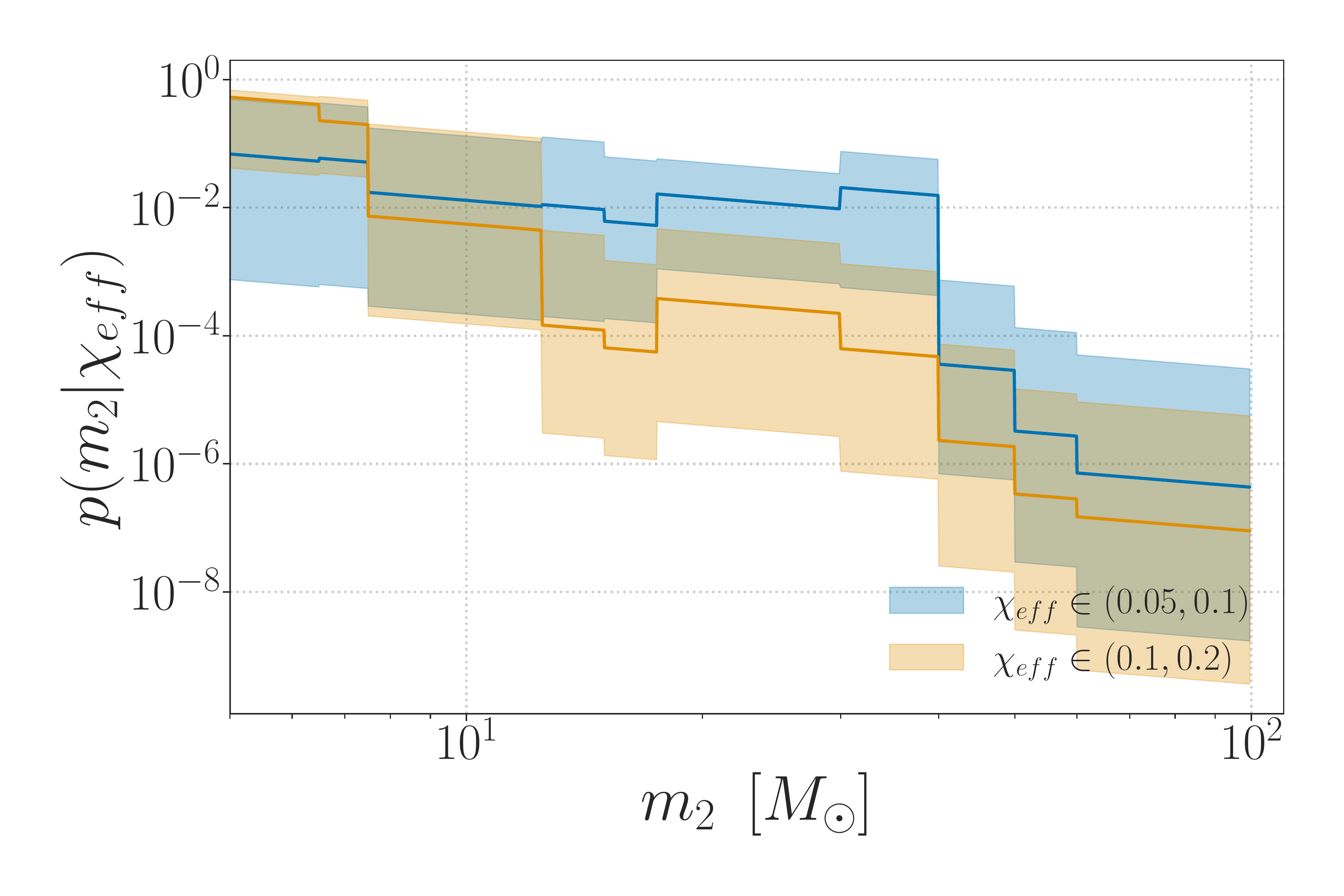}
\includegraphics[width=0.32\textwidth]{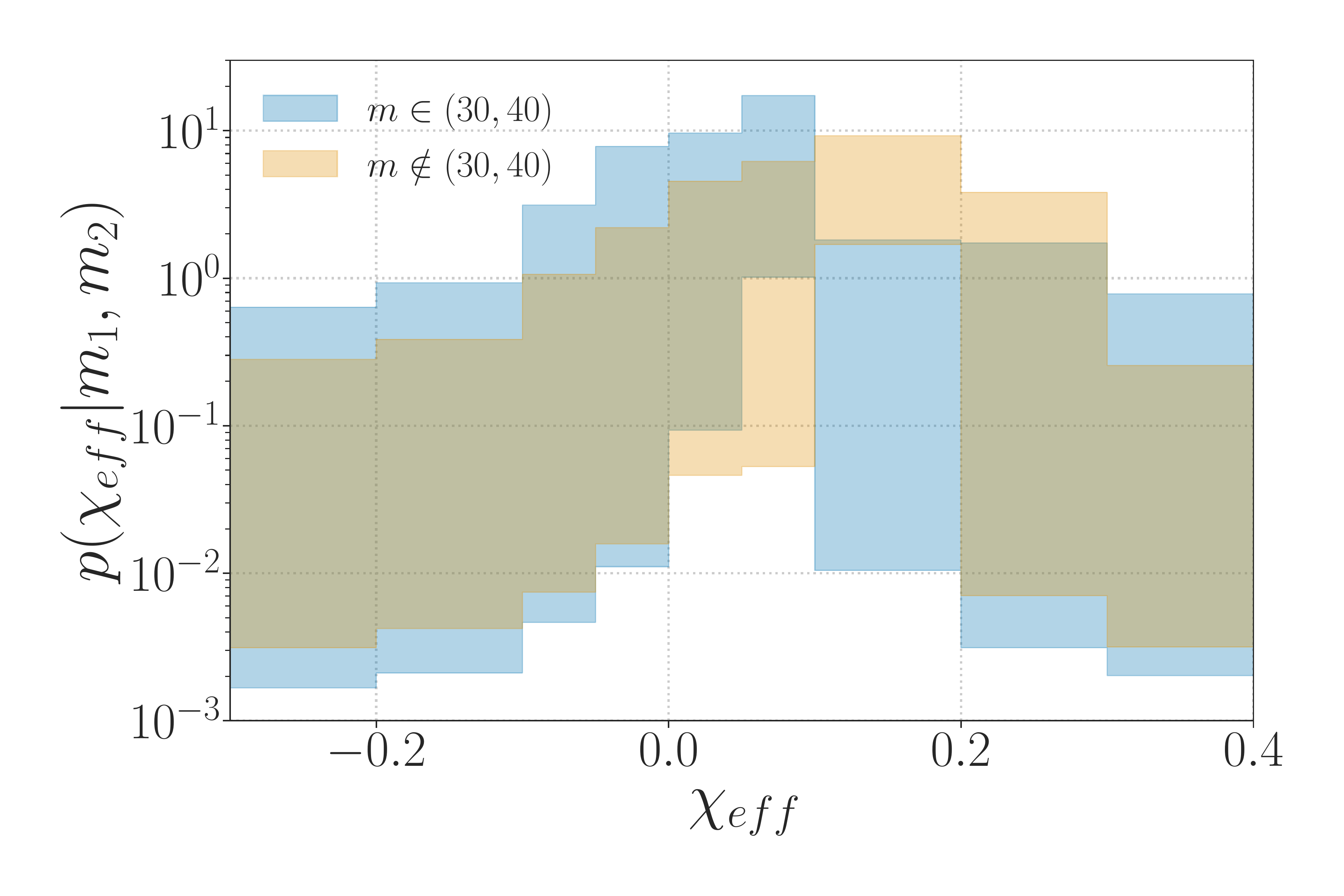}
\caption{\label{fig:kappa44} GWTC-3 results for $\kappa=4.4$}
\end{center}
\end{figure*}

In figure~\ref{fig:weights-vts-vs-kappa}, we first show how the posterior weights~($w^{\gamma}_i$) and detectable time-volumes~($\rm{VT}^{\gamma}_i$) of Eq.~\eqref{likelihood} required to compute our population likelihood vary with $\kappa$. In Figures.~\ref{fig:kappa09}, and \ref{fig:kappa44}, we show that the trends we find in the astrophysical mass and spin distributions of GWTC-3 BBHs are robust against variations in $\kappa$ with the upper and lower bounds of its measured values that were reported by \cite{KAGRA:2021duu}. We therefore conclude that our conclusions regarding the mass-spin sub-populations are not biased by our restrictions on $\kappa$.

\section{Variation of binning choices}
\label{sec:binning}
In this appendix, we show that the trends we find in the mass-spin population of GWTC-3 BBHs are robust against changes in choices of binning, for both masses and effective inspiral spins. We change both the location and number of bins which is done in two ways. In the first case we change the mass bins to while keeping the same $\chi_{\rm{eff}}$ bins as in Sec.~\ref{sec:results}. In the second case we change the $\chi_{\rm{eff}}$ bins while keeping the mass bins the same as they were in Sec.~\ref{sec:results}.

\begin{figure*}[htt]
\begin{center}
\includegraphics[width=0.48\textwidth]{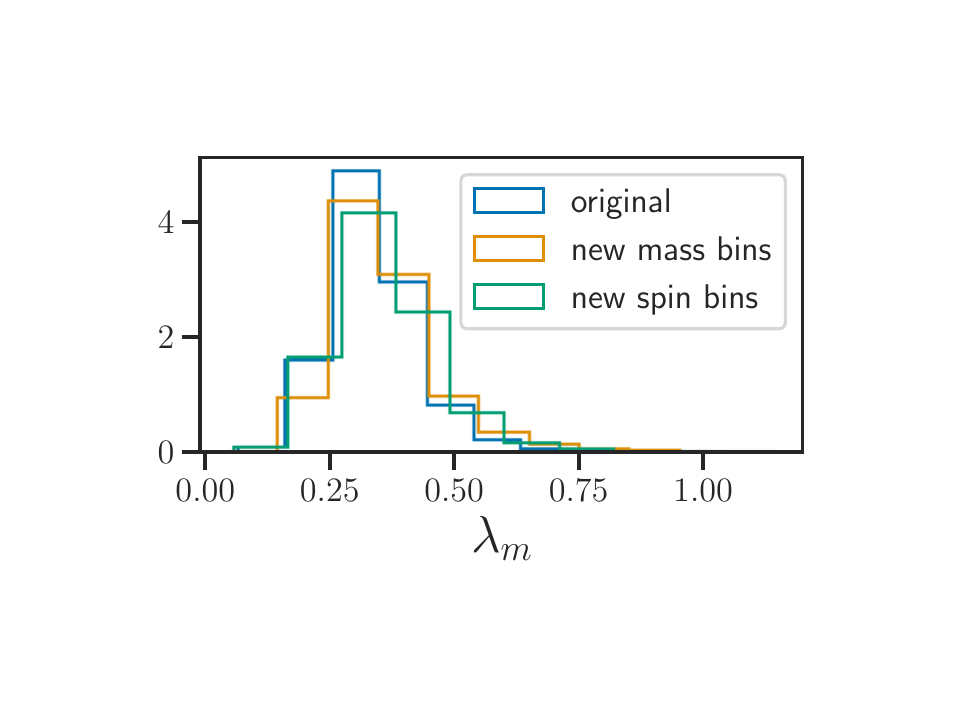}
\includegraphics[width=0.48\textwidth]{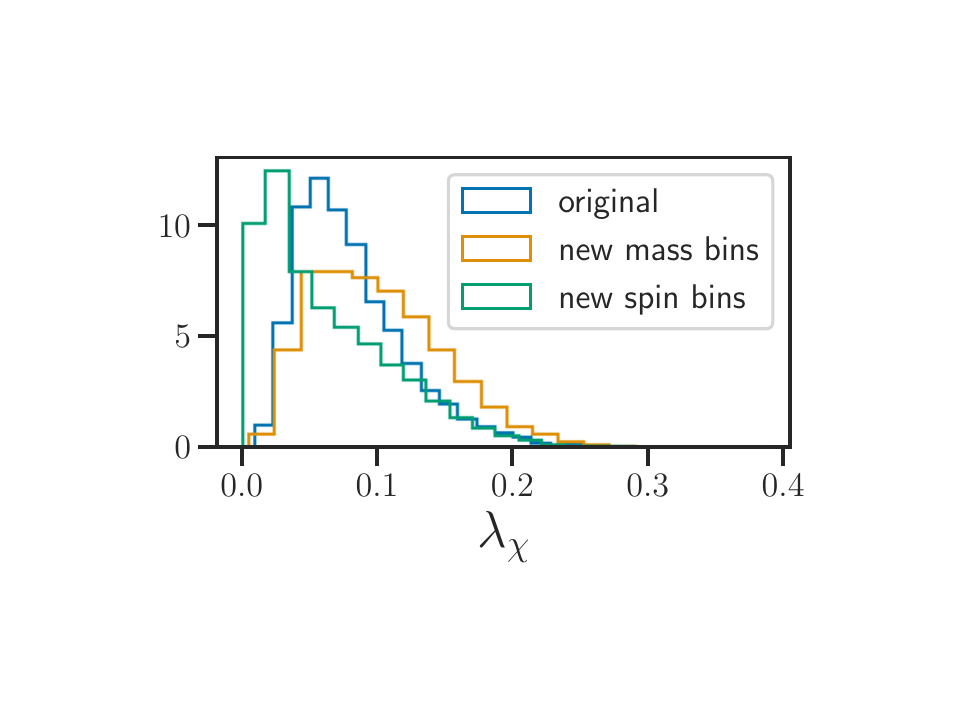}
\caption{\label{fig:ls-compare}Posterior distributions of the GP lengthscales along the mass~(per log mass bin center,\textit{left}) and spin~(per spin bin center \textit{right}) dimensions. }.
\end{center}
\end{figure*}

\begin{figure*}[htt]
\begin{center}
\includegraphics[width=0.32\textwidth]{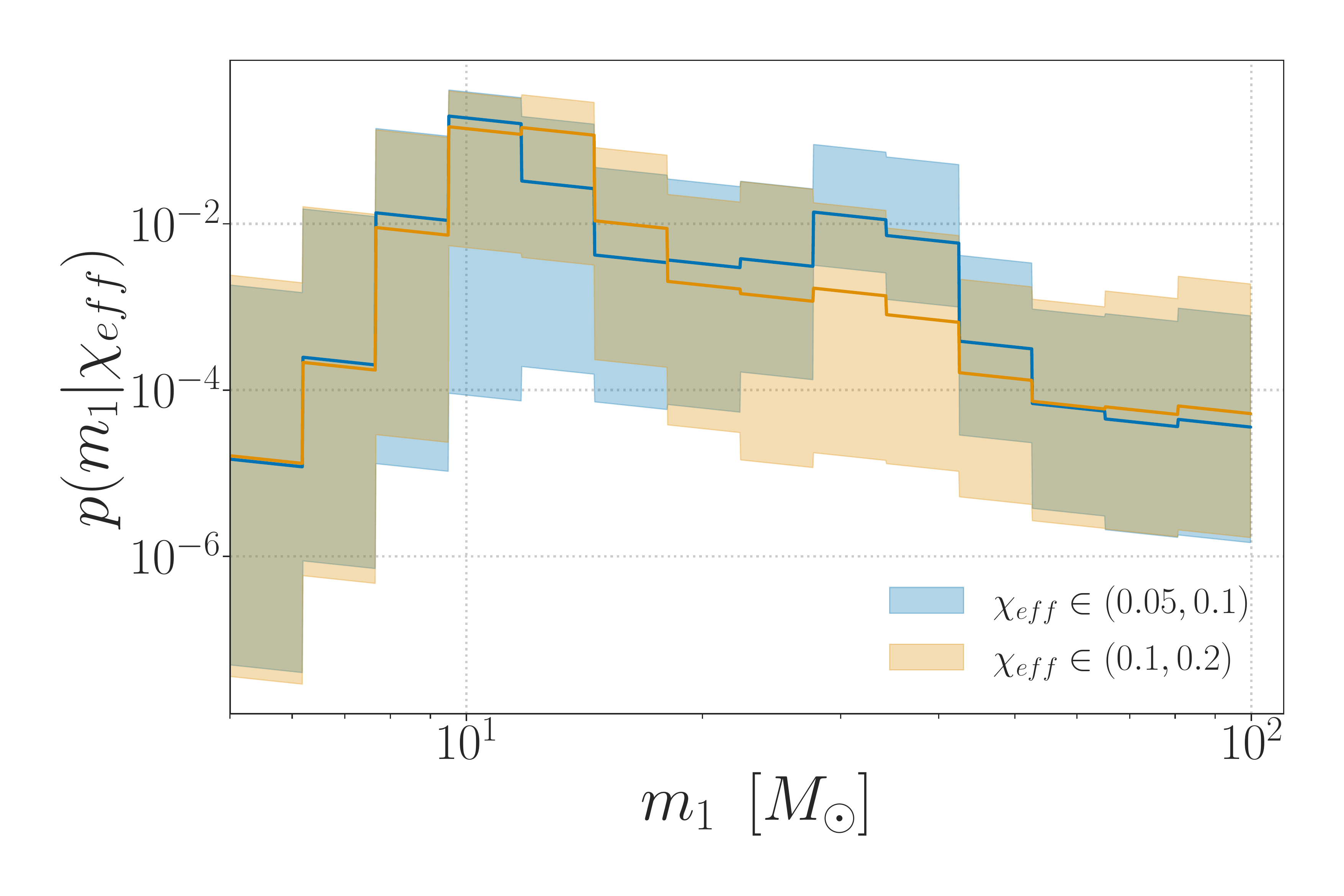}
\includegraphics[width=0.32\textwidth]{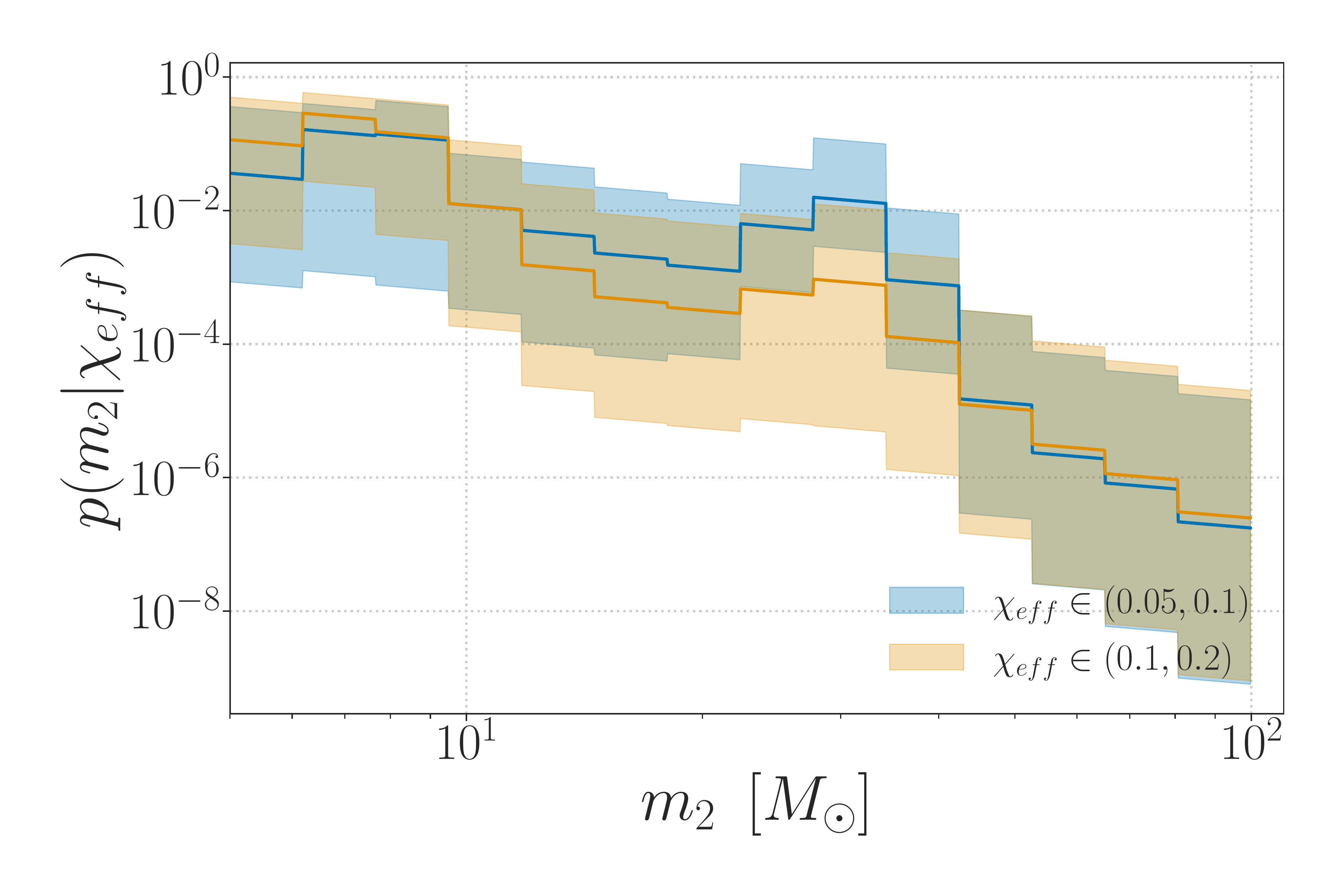}
\includegraphics[width=0.32\textwidth]{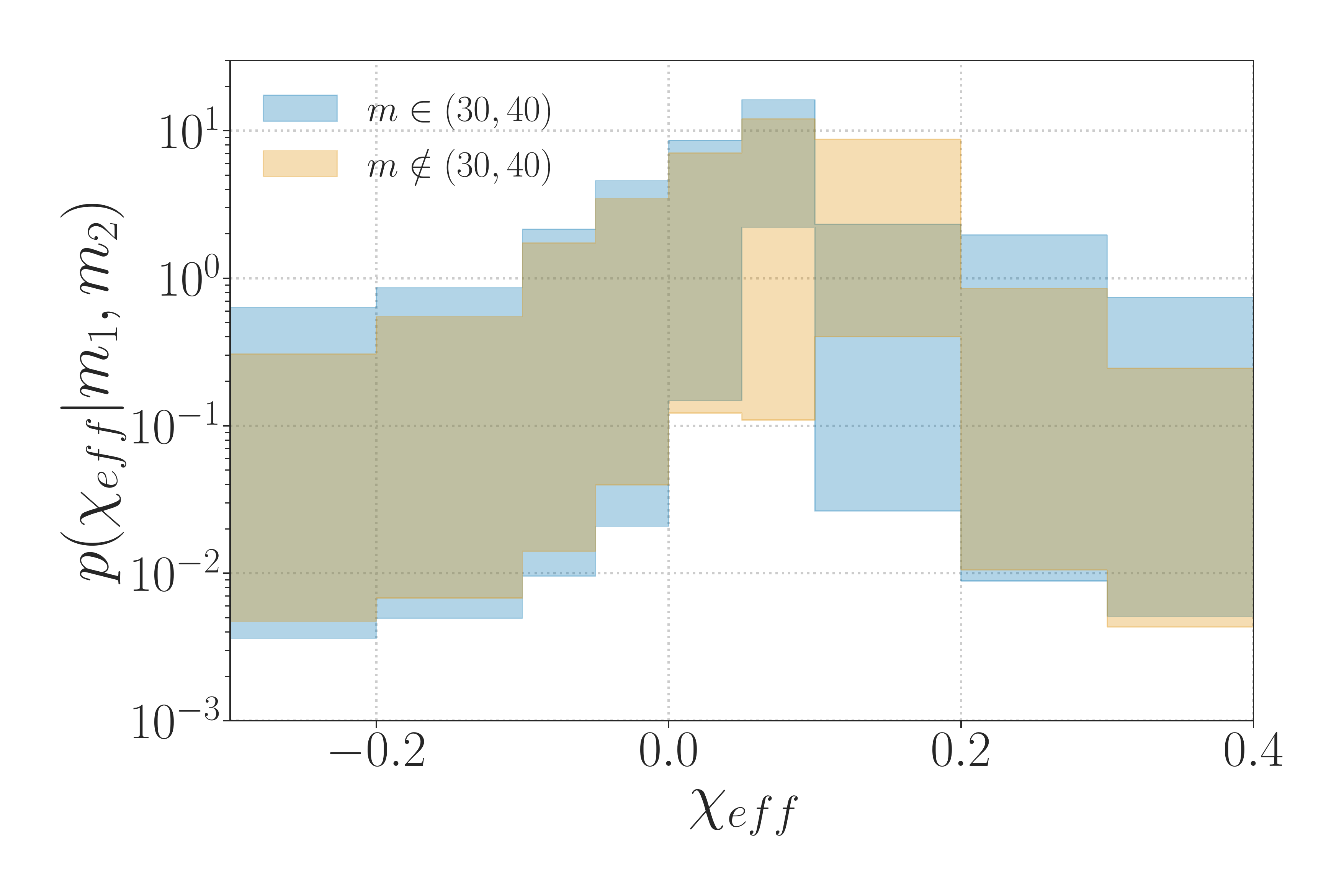}
\caption{\label{fig:change-mbins} GWTC-3 results for changed mass bins}
\end{center}
\end{figure*}

\begin{figure*}[htt]
\begin{center}
\includegraphics[width=0.32\textwidth]{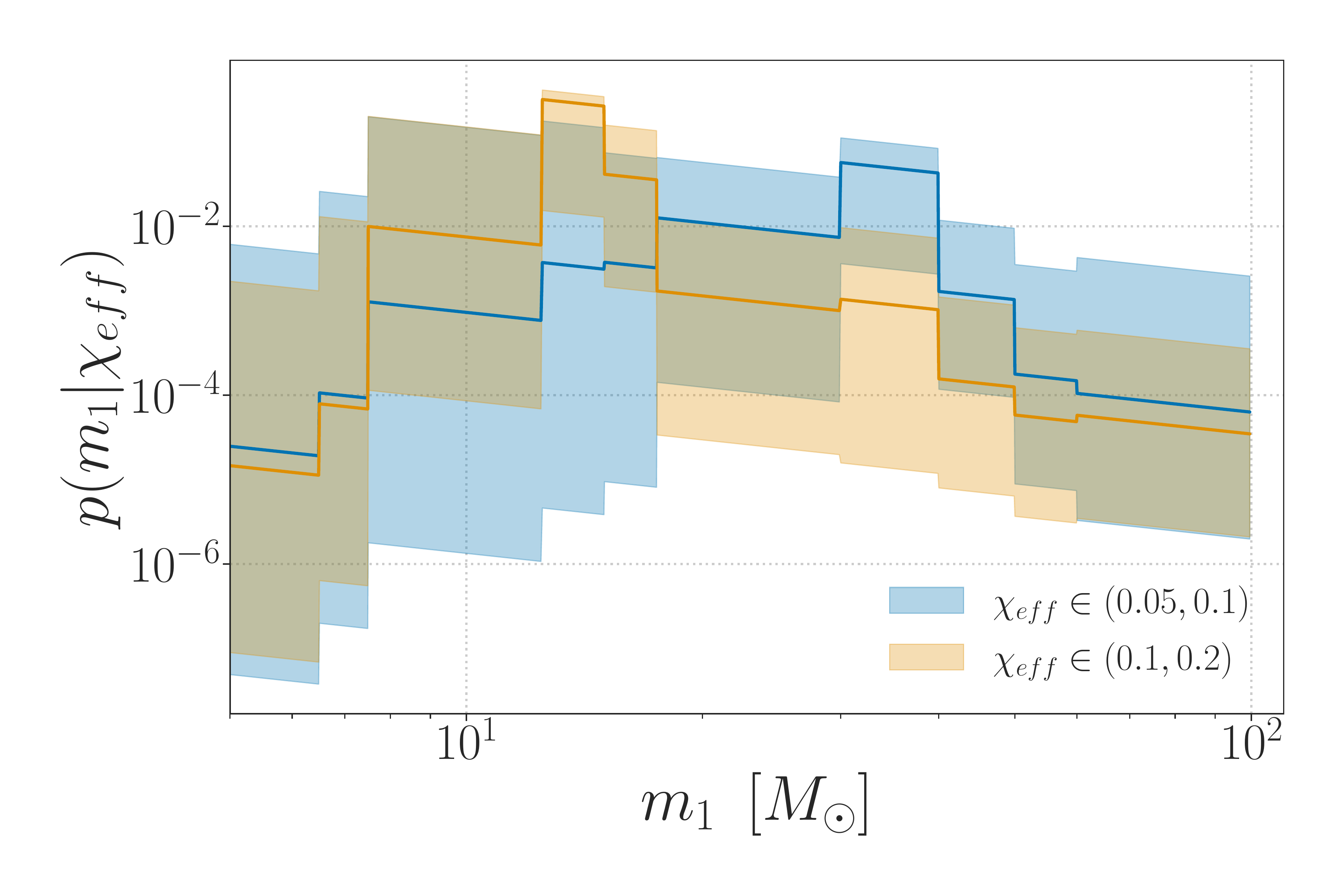}
\includegraphics[width=0.32\textwidth]{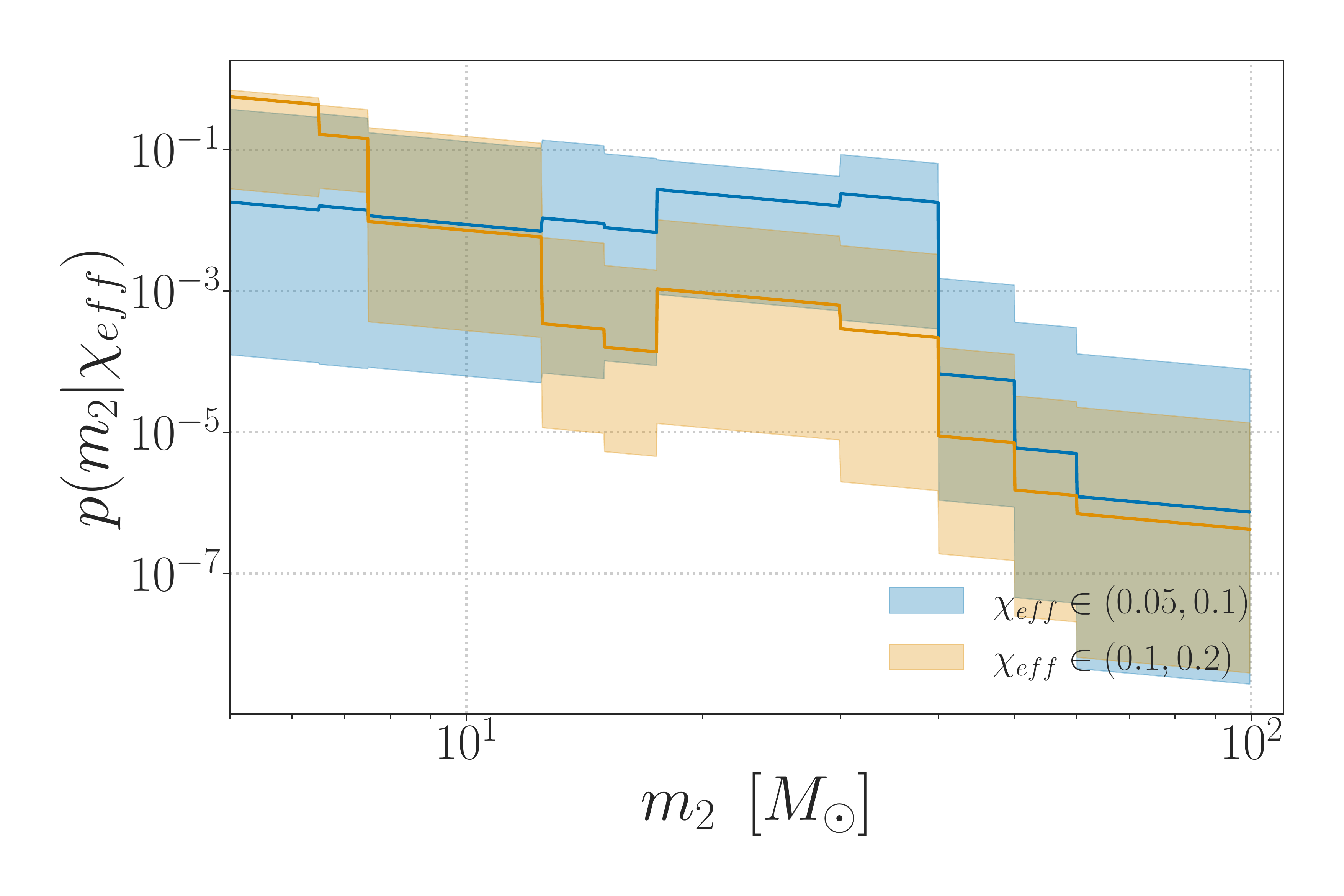}
\includegraphics[width=0.32\textwidth]{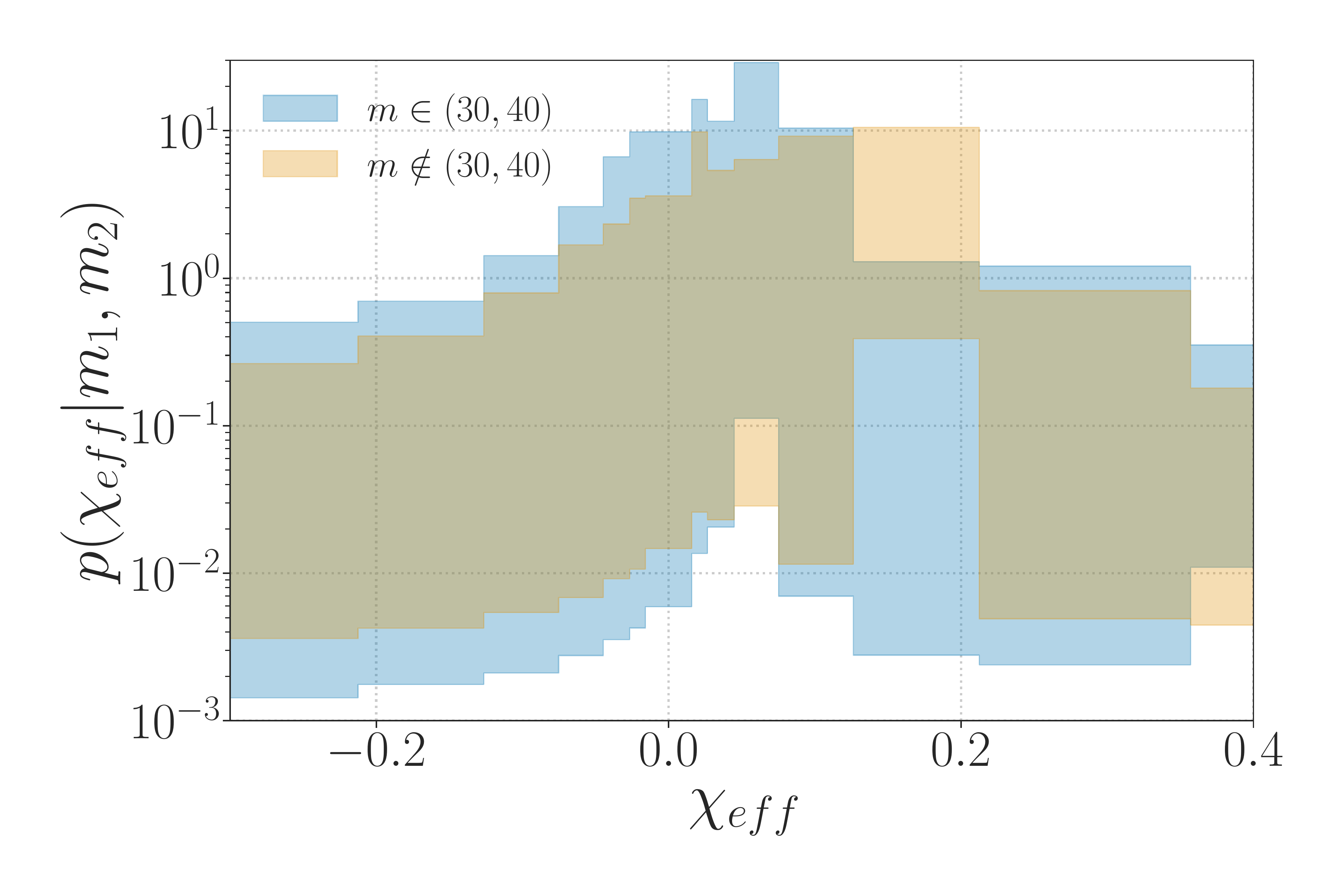}
\caption{\label{fig:change-chibins} GWTC-3 results for changed $\chi_{\rm{eff}}$ bins}
\end{center}
\end{figure*}

In both scenarios, we recover the same trends as in Sec.~\ref{sec:results}. This is expected given that the GP hyper-parameters such as the correlation lengths are themselves inferred from the data, with their priors determined by the distribution of bin centers~\citep{Ray:2023upk}. In Figure~\ref{fig:ls-compare} we show that the posterior distributions of the GP length scales along both the mass and spin dimensions remain consistent for all three choices of binning. This is indicative of the GP correlating more or less bins over the same range of parameter space depending on how many bins reside within that range, ultimately leading to consistency in the inferred shapes of the population disribution~\citep{Ray:2023upk}, as shown in Figures~\ref{fig:change-mbins} and \ref{fig:change-chibins}.

We note that increasing the number of bins makes the constraints broader, which is to be expected given that we are trying to fit a larger number of hyper-paramaters from the same dataset. Furthermore, the increased resolution does not reveal any additional features in the dataset which justifies our choice of binning used in Sec.~\ref{sec:results}.

\section{Scalability}
\label{sec:scalability}
Previous implementations of the BGP method had scalability issues which limited the resolution that could be achieved within a tractable inference run~\citep{Ray:2023upk}. This was mostly due to the cubic time-complexity of GP engines which in our case previously translated to $O(\prod_{\theta}N_{bins,\theta}^{3})$, where $N_{bins,\theta}$ are the number of bins along dimension $\theta$, and was bound to become prohibitive with the addition of more parameters or an increase in resolution. Specifically this particular form of complexity arises from the Cholesky decomposition of the GP's covariance kernel, which takes the following form:
\begin{eqnarray}
    K^{\gamma\gamma'}(\lambda,\sigma) &=& \sigma^2e^{-\sum_{\theta}\frac{(c^{\gamma}_{\theta}-c_{\theta}^{\gamma'})^2}{2\lambda_{\theta}^2}}\label{eq:cov1}\\
    \mathbf{K}&=& \mathbf{L}\mathbf{L}^{T},
\end{eqnarray}
 where $L$ is a lower triangular matrix. Given that $\gamma\in\{1,2,...,\prod_{\theta}N_{bins,\theta}\}$, and the cubic complexity of  Cholesky decomposition algorithms, the total computational cost is expected to scale as $O(\prod_{\theta}N_{bins,\theta}^{3})$. 

Fortunately, the structure of the kernel in Eq.~\eqref{eq:cov1} provides a nice work around to this problem. For example, the matrix in Eq.~\eqref{eq:cov1} can be expressed as a Kronecker product of smaller matrices (one for each BBH parameter):
\begin{eqnarray}
    \mathbf{K}&=&\bigotimes_{\theta}\mathbf{K}_{\theta}\\
    K_{\theta}^{\gamma\gamma'}&=& \sigma^{\frac{2}{n_{\theta}}}e^{-\frac{(c^{\gamma}_{\theta}-c_{\theta}^{\gamma'})^2}{2\lambda_{\theta}^2}},
\end{eqnarray}
where $n_{\theta}$ is the total number of BBH parameters considered in the population model. Exploiting the fact that the Cholesky decomposition of Kronecker product is the Kornecker product of Cholesky decompositions, as in:
\begin{eqnarray}
    \mathbf{L}&=&\bigotimes_{\theta}\mathbf{L}_{\theta}\\
    \mathbf{K}_{\theta}&=& \mathbf{L}_{\theta}\mathbf{L}^{T}_{\theta},
\end{eqnarray}
 we are able to implement an inference whose cost scales as $O(\sum_{\theta}N_{bins,\theta}^{3})$. This leads to higher dimensional generalizations of our models straight forward to implement in a scalable manner. Similarly, an increase in bin resolution along a particular dimension can also be implemented tractably, upto the scenario wherein we have around 50+ bins along each dimension, a level of resolution that is likely unnecessary for contemporary and near-future datasets.
%\section{Uncorrelated inference}

\bibliography{sample631}{}
\bibliographystyle{aasjournal}

\end{document}